\pdfoutput=1
\documentclass[english,aps,prb,twocolumn]{revtex4}
\usepackage{graphicx}
\usepackage[T1]{fontenc}
\usepackage[latin9]{inputenc}
\usepackage{textcomp}
\usepackage{amstext}

\begin{document}

\title{Ultrafast Photo-Induced Charge Transfer Unveiled by\\ Two-Dimensional
Electronic Spectroscopy}

\author{Oliver Bixner$^1$, Vladim\'{i}r Luke\v{s}$^2$, Tom\'{a}\v{s} Man\v{c}al$^3$, J\"{u}rgen~Hauer$^1$, Franz~Milota$^1$, Michael~Fischer$^4$,\\ Igor~Pugliesi$^5$, Maximilian~Bradler$^5$, Walther~Schmid$^4$, Eberhard~Riedle$^5$,
Harald~F.~Kauffmann$^{1,6}$,\\ and Niklas~Christensson$^1$}

\affiliation{$^1$Faculty of Physics, University of Vienna,
Strudlhofgasse 4, 1090 Vienna, Austria\\
$^2$Department of Chemical Physics, Slovak Technical
University, Radlinsk\'{e}ho 9, 81237 Bratislava, Slovakia\\
$^3$Institute of Physics, Faculty of Mathematics and
Physics, Charles University, Ke Karlovu 5, Prague 121 16, Czech
Republic\\
$^4$Department of Organic Chemistry, University of Vienna,
W\"{a}hringer Straße 38, 1090 Vienna, Austria\\
$^5$Lehrstuhl f\"{u}r BioMolekulare Optik,
Ludwig-Maximilians-University, Oettingenstrasse 67, 80538 Munich,
Germany\\
$^6$Ultrafast Dynamics Group, Faculty of Physics, Vienna University
of Technology, Wiedner Hauptstrasse 8 - 10, 1040 Vienna, Austria}

\begin{abstract}
The interaction of exciton and charge transfer (CT) states plays a
central role in photo-induced CT processes in chemistry, biology and
physics. In this work, we use a combination of two-dimensional
electronic spectroscopy (2D-ES), pump-probe measurements and quantum
chemistry to investigate the ultrafast CT dynamics in a lutetium
bisphthalocyanine dimer in different oxidation states. It is found
that in the anionic form, the combination of strong CT-exciton
interaction and electronic asymmetry induced by a counter-ion
enables CT between the two macrocycles of the complex on a 30~fs
timescale. Following optical excitation, a chain of electron and
hole transfer steps gives rise to characteristic cross-peak dynamics
in the electronic 2D spectra, and we monitor how the excited state
charge density ultimately localizes on the macrocycle closest to the
counter-ion within 100~fs. A comparison with the dynamics in the
radical species further elucidates how CT states modulate the
electronic structure and tune fs-reaction dynamics. Our experiments
demonstrate the unique capability of 2D-ES in combination with other
methods to decipher ultrafast CT dynamics.
\end{abstract}

\maketitle

\section{Introduction}

Charge transfer (CT) is a ubiquitous process in nature representing
the initial step in many chemical reactions and bioenergetic
pathways.\cite{Marcus1985} Photo-induced CT is intimately related
to harnessing and conversion of radiative energy in photovoltaics
\cite{Deibel2010,Durrant2010,Herrmann2011}, photosynthesis
\cite{Blankenship2002,Meech1986,Novoderezhkin2007} and catalysis
\cite{Bauer2005}. CT states, representing an excitation in a
molecular aggregate where the electron and hole are located on
different chromophores, are characterized by low transition dipole
moments and strong sensitivity to environmental perturbations.
\cite{Renger2004,Mancal2006} Due to their low transition dipole
moments, CT states participate in photo-induced reactions
via the interaction with optically allowed (bright) states. If the
interaction with the CT states is weak, the resonance interaction
between locally excited (LE) states leads to excited states which
are delocalized over the chromophores in the aggregate (exciton
states). For a more pronounced CT interaction, the mixing of
excitonic and CT states leads to (delocalized) eigenstates
exhibiting various degrees of charge separation. After
photo-excitation, interaction with the nuclear degrees of freedom
enables relaxation within the manifold of excited states, ultimately
leading to population of the CT states.

The influence of the CT states on the electronic structure depends on the relative magnitude of resonance- (Coulombic interaction between LE states) and CT-coupling (between LE and CT states, depending on the wave function overlap \cite{MayKuhn2000}).
This elementary interplay can be studied in model systems where the
different couplings can be controlled via the inter-pigment
distance, and where the energies of the CT states can be altered
electrochemically. One such class of molecules is
bisphthalocyanines, which comprise two macrocyclic phthalocyanine
moieties held together at a fixed distance by a rare earth cation.
\cite{Koike1996,Moussavi1988}
The close proximity of the two rings (2.8 {\AA} for lutetium) gives rise to delocalized charge distributions and results in strong CT coupling due to significant wavefunction overlap.\cite{Ishikawa1992} The resulting high degree of charge carrier
mobility readily explains their semiconductivity\cite{Bouvet1990,Maitrot1987} and electrochromic behavior \cite{Kadish2001,Nicholson1982}.

\begin{figure}
\includegraphics[width=8.6cm]{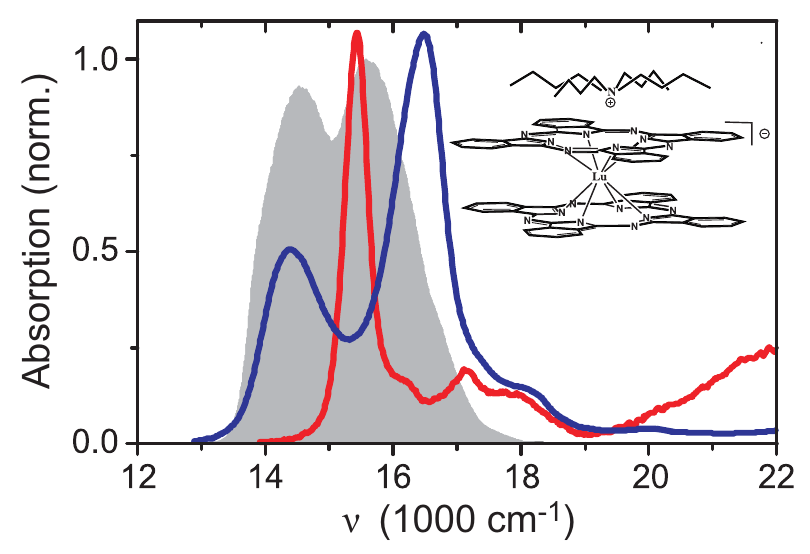}
\caption{\label{fig:Fig01}Linear absorption of
[LuPc$_{2}$]$^{-}$TBA$^{+}$ in benzonitrile (blue line) and
[LuPc$_{2}$]$^{\bullet}$ in toluene (red line) shown together
with the pulse spectrum used in the 2D experiments (grey). The inset
shows the structure of [LuPc$_{2}$]$^{-}$ and the position of
the TBA$^{+}$ counter-ion.}
\end{figure}

In its electronic ground state, lutetium bisphthalocyanine
([LuPc$_{2}$]) is a stable radical ([LuPc$_{2}$]$^{\bullet}$)
with an intermediate oxidation state of $-$1.5 on each ring
\cite{Ostendorp1996}. [LuPc$_{2}$]$^{\bullet}$ cannot stabilize
inter-ring CT states, and the electronic structure can be understood
via resonance interaction between the LE states. The electronic structure
will be similar to that of an H-type dimer, which gives rise to a single
transition in the linear absorption spectrum (Fig.~\ref{fig:Fig01}).
On the other hand, the anion ([LuPc$_{2}$]$^-$)
exhibits CT states in the same energy range as the LE states.\cite{Ishikawa1992} The
additional CT coupling drastically changes the electronic structure
and gives rise to two bands in the linear absorption spectrum
(Fig.~\ref{fig:Fig01}). Therefore, the comparison of [LuPc$_{2}$] in
different oxidation states provides a perfect opportunity to
pinpoint the influence of the CT states on the electronic structure
and dynamics.

The structural and electronic similarities favor [LuPc$_{2}$]$^{-}$ as
a biomimetic model system for studies of the special pair in the
photosynthetic reaction center (RC) \cite{Blankenship2002}. However,
for isolated [LuPc$_{2}$]$^{-}$, the CT states representing forward
and backward CT between the rings are energetically degenerate, and
the electronic structure can be described by charge resonance
states.\cite{Ishikawa1992} The presence of an environmental asymmetry
breaks the energetic degeneracy of the CT states, lowering the energy
of the CT state on one side of the complex and enables a net transfer
of charge after photo-excitation. In the RC, such asymmetry arises from
differences in the electronic coupling among pigments due to slightly
different spacings between the cofactor units \cite{Kolbasov2000},
distinct dielectric environments due to intra-protein electric fields \cite{Gunner1996,Wakeham2005,Steffen1994}, and different hydrogen
bonding pattern around the central units \cite{Moore1999}. In other
redox systems, the asymmetry may be provided by counter-ions. The
formation of ion associates can have profound influence on the
electronic structure and the path of chemical reactions.\cite{Cembran2005,Vakarin2002,Marcus1998}
For weak CT coupling (non-adiabatic CT), the role of the counter-ion
has been studied, and it has been shown that the counter-ion
can actively or passively control the rate of CT depending on the strength of
association.\cite{Vakarin2002,Piotrowiak1994,Piotrowiak1993,Marcus1998}

Conventionally, CT has been studied in systems where the CT coupling is larger than the resonance coupling, implying that CT proceeds without interference from energy transfer processes. However, in systems like the natural RCs or in novel architecture for artificial light harvesting, the two couplings can be of equal magnitude and both energy transfer and CT will take place simultaneously. The interplay of resonance- and CT-coupling in these systems tunes the excited state evolution and the dynamics of the charge separation and energy transfer processes. Such excited state dynamics gives rise to complicated spectral signatures, making a detailed investigation of the underlying mechanisms with conventional techniques difficult. In this work we employ two-dimensional- (2D) and ultra-broadband pump-probe-spectroscopy to disentangle the excited state dynamics in a metal bridged dimer ([LuPc$_{2}$]), where the resonance- and CT-couplings are of equal magnitude. The influence of CT states and CT coupling on the excited state structure and dynamics is further studied via electrochemical tuning of the CT state energies. In addition, we investigate how specific details of ion-pairing enable ultrafast charge transfer across the complex. The combination of various approaches, both theoretical (quantum chemistry and density matrix propagation) and experimental (absorption, pump-probe and 2D spectroscopy), gives a detailed picture of the ultrafast CT dynamics.

\section{Experimental Methods}

Lutetium(III)bisphthalocyanine-tetrabutyl-ammonium salt
([LuPc$_{2}$]$^{-}$TBA$^{+}$), and its radical form
[LuPc$_{2}$]$^{\bullet}$ were synthesized via a base catalyzed
cyclotetramerization reaction of phthalonitrile with lutetiumacetate
in the bulk-phase at elevated temperature. The synthesis was carried
out using known procedures from the literature.\cite{Moussavi1988,Konami1989}
Both the synthetic preparation and
the characterization of the complex are described in detail in the
Supporting Information.\cite{SuppMaterial}

The setup and methodology used for the 2D experiments has been described in
detail previously.\cite{MilotaKauffmannI2009} In the experiments
reported here, a home built NOPA operated at 200 kHz repetition rate
generated pulses centered at 15000 cm$^{-1}$ with a bandwidth of 2500
cm$^{-1}$ FWHM (Fig.~\ref{fig:Fig01}). The FWHM of the intensity
autocorrelation was 13 fs (9.2 fs pulse duration), while the
corresponding width assuming a flat phase over the pulse spectrum
was 10 fs. Two phase stable pulse pairs were generated by
diffracting the NOPA pulses through a transmission grating. In the
2D experiment, the delay between the first two pulses, $t_{1}$, was
scanned with interferometric accuracy, while the delay between the
second and third interaction, $t_{2}$, was held constant. The
signal radiated in the phasematching direction was overlapped with
the local oscillator (LO), and the signal field was recovered by
spectral interferometry. The 2D spectrum as a function of $v_{1}$ and
$v_{3}$ was obtained by Fourier transforming over $t_{1}$ and
addition of rephasing and non-rephasing signal contributions. The
absolute phase of each 2D spectrum was determined by adjusting the
LO delay and phase to optimize the overlap of the projection of the
real part of the 2D spectrum with the spectrally resolved pump-probe
signal.\cite{SuppMaterial} [LuPc$_{2}$]$^{-}$TBA$^+$
dissolved in benzonitrile was used in a wire-guided flow jet giving
a film thickness of 230 $\mu$m. The structure and linear absorption
spectrum of [LuPc$_{2}$]$^{-}$TBA$^{+}$ are shown in Fig.~\ref{fig:Fig01} together with the NOPA spectrum used in the
experiments.

Broadband pump-probe measurements were carried out in a
setup described in detail previously.\cite{Megerle2009} To extend
the probe range to the NIR region, a home built OPA operating at
8300 cm$^{-1}$ was used to generate white-light from 9000-25000
cm$^{-1}$.\cite{Bradler2009,Herrmann2011} To cover the bands in the
UV region (25000-32000 cm$^{-1}$), white-light was generated by
focusing the output from the Ti:Sapphire laser into a CaF$_2$ plate.
\cite{Megerle2009} Pump pulses of approximately 40 fs duration were
used to selectively excite the two bands at 14200 and 16200
cm$^{-1}$ giving a time resolution of 50 fs.

\section{Results}
\subsection{Experimental Results}

The interplay of exciton and CT states and their influence on the
excited state dynamics in [LuPc$_{2}$]$^{-}$ have been studied
previously with pump-probe- and two-color photon echo-spectroscopy.
\cite{Prall2005,PrallIshikawa2004} The pump-probe measurements
revealed that the transient spectra were independent of which of the
two bands was pumped, indicating rapid relaxation in the excited state.
To resolve the excited state dynamics we turned to
2D spectroscopy.\cite{Jonas2003} By correlating the optical
coherence evolution between the first and third time interval in a
four-wave mixing sequence, 2D spectroscopy maps the system dynamics
onto two dimensions, thereby minimizing spectral congestion and
provides a very detailed picture of the system\textquoteright{}s
evolution.\cite{Cho2008}

\begin{figure*}
\includegraphics[width=17.2cm]{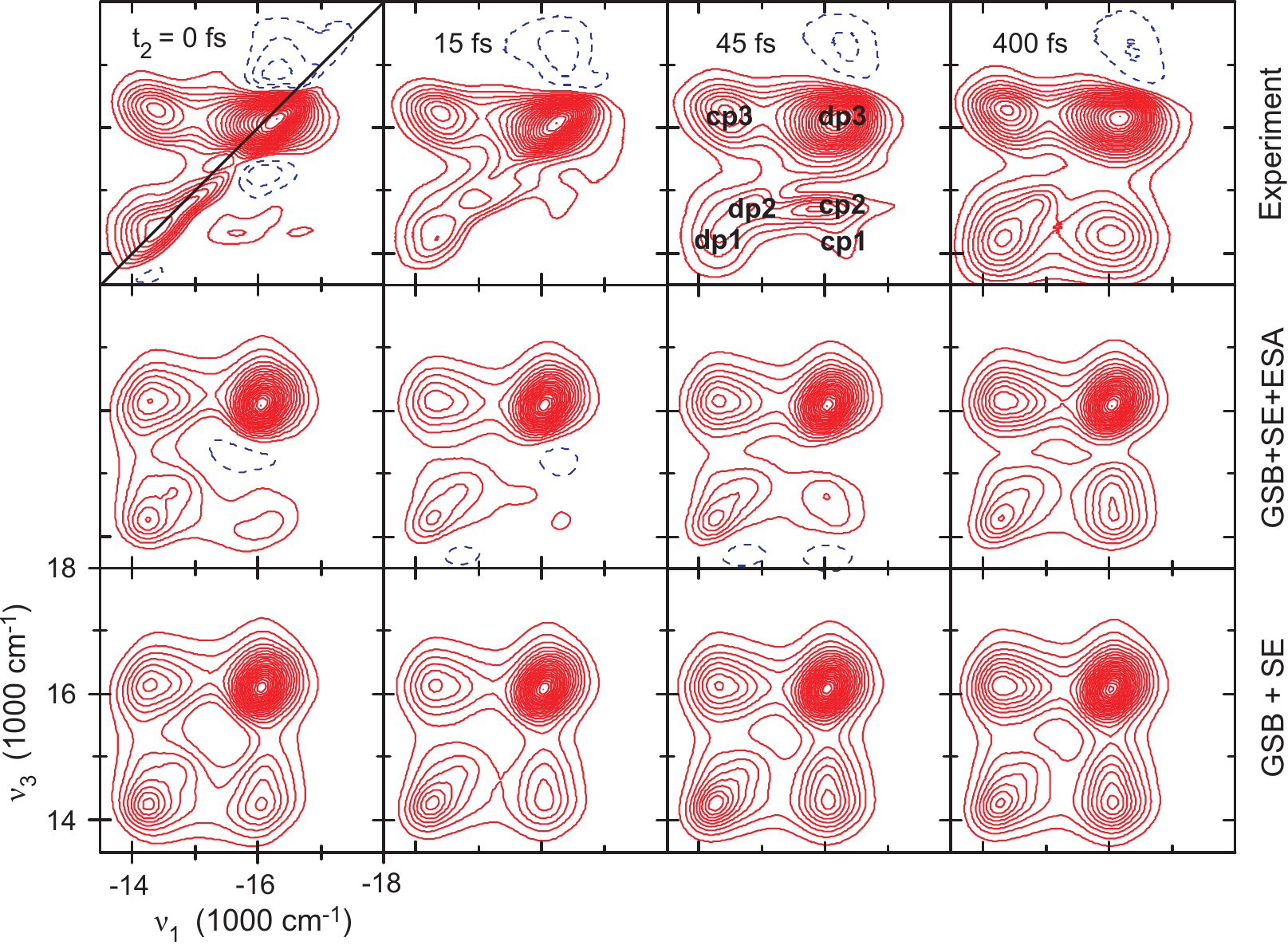}
\caption{\label{fig:Fig02}2D spectra of [LuPc$_{2}$]$^{-}$TBA$^{+}$ at different waiting times
($t{}_{2}$) indicated in the Figure. The top row shows the
experimental results, while the middle and bottom rows show the
simulated total spectra and ground state bleach (GSB) plus
stimulated emission (SE) contributions, respectively. All spectra
have been normalized to the maximum of the total signal, and contour
lines are drawn in 5 \% increments starting at 5 \%. Positive signal
contributions (GSB+SE) are shown in red while negative (ESA) are
shown in blue dash. The points used for the kinetics in
Fig.~\ref{fig:Fig03} are shown for $t_{2}=$ 45 fs.}
\end{figure*}

Fig.~\ref{fig:Fig02} shows the real part of the 2D spectrum of
[LuPc$_2$]$^{-}$TBA$^+$ in benzonitrile for $t_{2}=$ 0, 15, 45, and 400
fs. The waiting time $t_{2}$ denotes the delay between the first
and the second pulse pair, and corresponds to the probe delay in a
pump-probe experiment. All spectra have been normalized to their respective maxima, and
contour lines are drawn in 5 \% increments. In Fig.~\ref{fig:Fig02}, positive signals correspond to stimulated emission (SE) and ground state bleach (GSB), while excited state absorption (ESA) results in negative signals. The 2D spectrum at $t_{2}=$ 15 fs
(outside pulse overlap) shows two diagonal peaks
corresponding to the two bands in the linear absorption spectrum. A
closer inspection of the low energy diagonal peak reveals that it is more accurately described as two separate peaks at
$v_{3}=$ 14200 and 14800 cm$^{-1}$, respectively. Above the high
energy diagonal peak we observe small amounts of ESA (blue dashed
lines). In addition to the diagonal peaks and ESA contributions, one
cross-peak at $v_{1}$, $v_{3}=$ 14200, 16200 cm$^{-1}$ is clearly
visible. This cross-peak shows that the transitions responsible for
the main bands in the linear absorption spectrum share a common
ground state. Surprisingly, there is no corresponding cross-peak
below the diagonal. For longer $t_{2}$ we find that the
cross-peak below the diagonal recovers. However, the recovery of the
cross-peak is not uniform. The 2D spectrum at $t{}_{2}=$ 45 fs shows
that the cross-peak at $v_{3}=$ 14800 cm$^{-1}$ grows much faster
than the one at 14200 cm$^{-1}$. For longer $t_{2}$, the cross-peak
below the diagonal becomes more symmetric, but its amplitude remains
weaker than the amplitude of the opposite cross-peak for all
$t_{2}$.

\begin{figure}
\includegraphics[width=8.6cm]{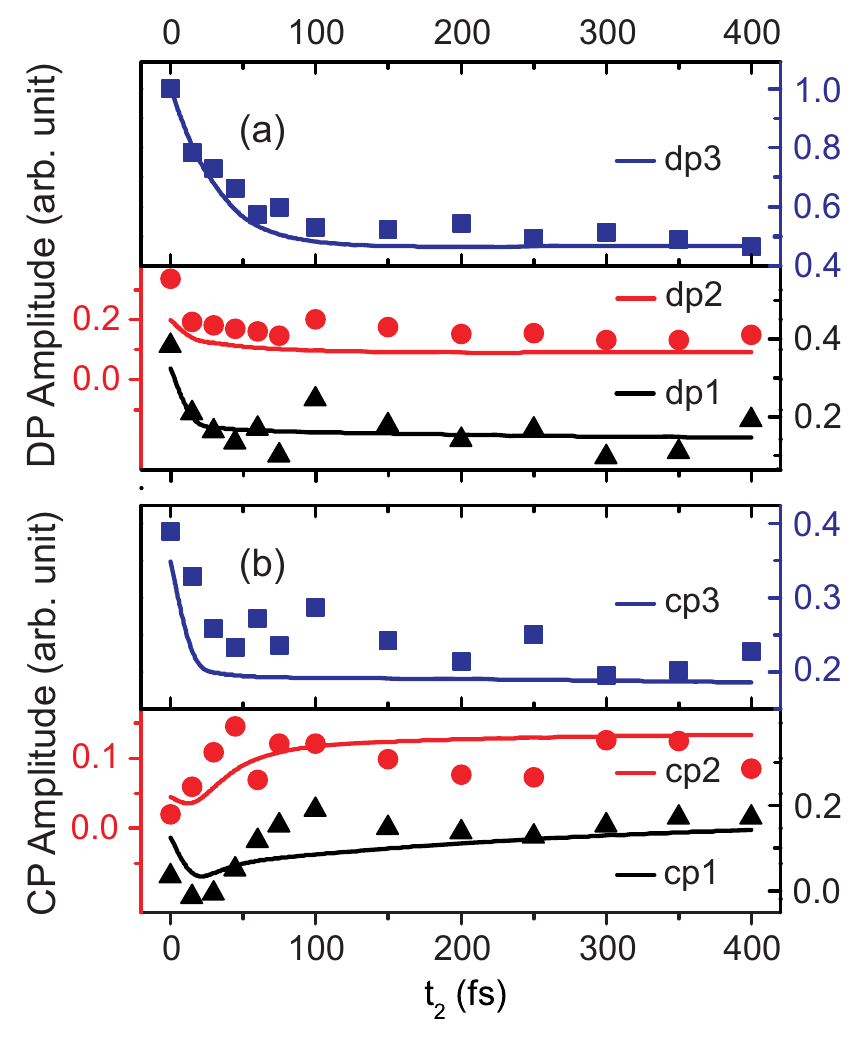}
\caption{\label{fig:Fig03}Kinetics of the different points in the 2D
spectra indicated in Fig.~\ref{fig:Fig02}. (a) Diagonal peak kinetics for
dp1 (black triangles), dp2 (red circles) and dp3 (blue squares,
upper panel) together with the simulations (solid lines in the same
color). The different curves have their own y-axis as indicated by
the colors. (b) Cross-peak kinetics for cp1 (black triangles), cp2
(red circles) and cp3 (blue squares, upper panel) together with the
simulated curves (solid lines).}
\end{figure}

The 2D spectra in Fig.~\ref{fig:Fig02} do not provide an accurate picture
of the dynamics in the system because of the normalization. The
variation of the amplitudes of the different peaks as a function of
$t{}_{2}$ provides information on population dynamics of the system
which cannot be obtained from the normalized spectra. To analyze the
kinetics, we show in Fig.~\ref{fig:Fig03} the time evolution of the
volume of a 30$\times$30 cm$^{-1}$ box for the diagonal peaks at
14200 (dp1), 14800 (dp2) and 16200 cm$^{-1}$ (dp3), and the
cross-peaks $v_{1}$, $v_{3}=$ 16200, 14300 cm$^{-1}$ (cp1), $v_{1}$,
$v_{3}=$ 16200, 14800 cm$^{-1}$ (cp2), and $v_{1}$, $v_{3}=$ 14200,
16200 cm$^{-1}$ (cp3). The positions of these points are also shown
in Fig.~\ref{fig:Fig02} for $t_{2}=$ 45 fs. The kinetics of the diagonal
peaks all display a fast decay to about 50 \% of the
value at $t_{2}=$ 0. The amplitude of dp3 decays on a 30 fs
timescale, while dp1 and dp2 show an almost constant amplitude
outside pulse overlap ($t_{2}\geq$ 15~fs). The cross-peaks exhibit
more complicated behavior. While cp3 behaves qualitatively in the
same way as dp3, the cross-peaks below the diagonal show distinct
dynamics. The amplitude of cp2 rises on a 20 fs timescale followed
by a quick decay. After about 50 fs, the rise is complete and the
signal shows a slow oscillation around a stationary value. The
amplitude of cp1 shows a delayed rise, reaching its peak value after
about 100 fs. After reaching their final values, cp1 and cp2 both
oscillate with a frequency of 160 cm$^{-1}$. The same modulation is
found in pump-probe \cite{Prall2005}, two-color photon echo
\cite{PrallIshikawa2004}. We can assign this mode, based on our quantum chemical
calculations, to a modulation of the Lu-N distance.\cite{SuppMaterial}

The appearance of the 2D spectrum for $t_{2}\geq$ 30~fs deviates from
what we expect for a coupled dimer, insofar as it does not display a
characteristic four peak pattern \cite{KjellbergPulleritsI2006}. The
absence of a cross-peak below the diagonal gives rise to a
pronounced asymmetry of the spectra. This effect clearly remains
outside pulse overlap, but diminishes as the population relaxes from
the upper band (decay of dp3). This points to ESA from the upper band
as the origin of the cancellation of the cross-peak below the
diagonal. As the ESA signal decays, the underlying GSB contribution
becomes visible. Such relaxation should give rise to a SE signal
in the region of the cross-peak below the diagonal (relaxation cross-peak).
For this reason, we expect this cross-peak to be stronger than the corresponding one above the diagonal (cp3) when relaxation from the upper band has been completed.
However, we find that the cross-peak below the diagonal remains
weaker for all $t_{2}$. This observation could in principle be explained by loss
of the SE signal due to relaxation to a state outside the spectral
window or into a state with low transition dipole moment. However,
we see no indication of such relaxation in the kinetics of either
dp1+dp2 or cp1+cp2. If the population remains in the lower band, the
positive SE signal could be hidden by a negative ESA
signal overlapping with the positive GSB of the low energy band.
While this would explain the 2D kinetics, it cannot explain the
reported lack of fluorescence from the low energy band
\cite{Prall2005}.

\begin{figure}
\includegraphics[width=8.6cm]{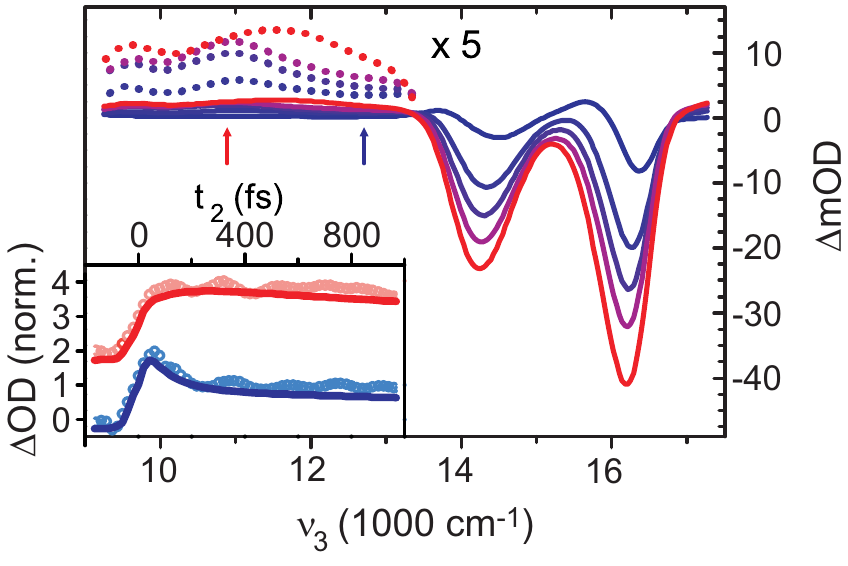}
\caption{\label{fig:Fig04}Pump-probe spectra of
[LuPc$_{2}$]$^{-}$TBA$^{+}$ in benzonitrile with a focus on the
NIR region for $t_{2}=$ 0.1, 0.6, 1.5, 3, and 10 ps (from red
to blue). The inset shows the (normalized) kinetics at the
frequencies indicated by the arrows in the main Figure, i.e. at
10870 cm$^{-1}$ (blue circles) and 12500 cm$^{-1}$ (red
circles). The solid lines are the corresponding fits. In pump-probe
spectroscopy, the standard sign convention (i.e. $\Delta$OD) is
opposite to that in 2D, e.g. ESA gives rise to positive signal
contributions. }
\end{figure}

It thus becomes apparent that the 2D experiment, despite the 4500 cm$^{-1}$ probe
range, is not able to follow all relaxation processes in the system
and cannot provide on its own the basis for a complete relaxation model. To be able
to probe the response of (possible) low energy
states in the NIR which are not covered in our 2D experiment, we
turned to broadband pump-probe spectroscopy. Fig.~\ref{fig:Fig04} shows
the pump-probe spectra in the NIR region for a few selected delays.
At $t_{2}=$ 100~fs, a broad ESA signal covers the entire region
from 9000-13000~cm$^{-1}$. For longer t$_2$, the ESA around 12500
cm$^{-1}$ (blue arrow) seems to decay significantly faster than the
ESA below 11000~cm$^{-1}$ (red arrow). The single wavelength fits
shown in the inset reveal that the ESA at 10870~cm$^{-1}$ decays
with a 3.8 ps exponential component, which is equal to the lifetime
of the lowest excited state obtained from a global fit.\cite{SuppMaterial}
On the other hand, the signal at 12500~cm$^{-1}$ shows
an additional 70-100 fs decay component. The kinetics in this
spectral region reflects the sum of ESA and SE contributions. The
additional fast decay of the signal at 12500 cm$^{-1}$ can thus be
interpreted as the delayed rise of SE from an almost dark state with
a transition in the NIR region as discussed above. The SE signal at
12500 cm$^{-1}$ is about 20 times weaker ($\sim$1 compared to 20
mOD) than the main GSB peak at 14200 cm$^{-1}$. The low transition
dipole moment of this transition readily explains the lack of
fluorescence in this system. For a 3.8 ps lifetime, we estimate a
fluorescence quantum yield in the 10$^{-5}$ range, beyond the
detection sensitivity of our and previous experiments
\cite{Prall2005}.

The discussion of 2D and pump-probe spectra above provides a rough
overview of electronic structure and ultrafast dynamics in
[LuPc$_{2}$]$^{-}$TBA$^+$. The states initially prepared by
pumping the upper band decay to the low energy bright band on a 30
fs timescale. The presence of two diagonal peaks, and the distinct
relaxation dynamics associated with cp1 and cp2 points to two
separate electronic states in this band (Fig.~\ref{fig:Fig03}). The
population of these two states flows on a 100 fs timescale into an
almost dark state with a transition in the NIR range
(Fig.~\ref{fig:Fig04}).

\begin{figure}
\includegraphics[width=8.6cm]{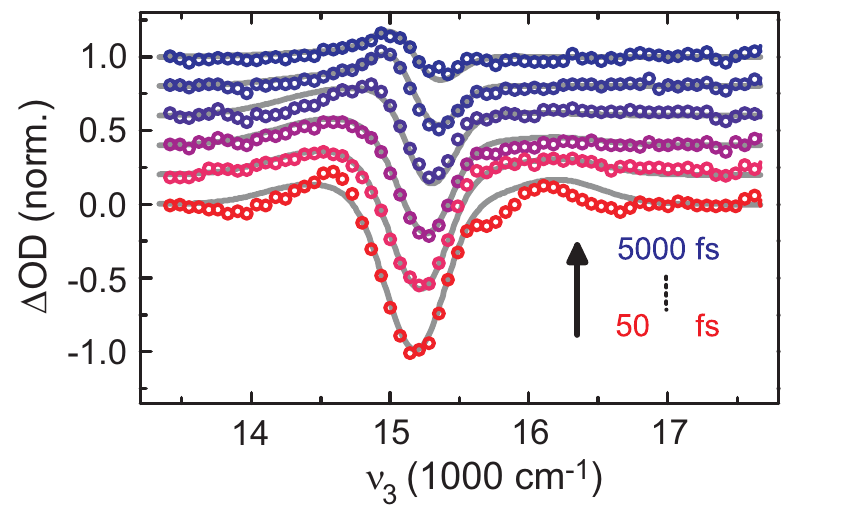}
\caption{\label{fig:Fig05}Pump-probe spectra of
[LuPc$_{2}$]$^{\bullet}$ in toluene for $t_{2}=$ 0.05, 0.2, 0.5,
1, 3, and 5 ps (from bottom to top). The spectra have been
shifted for ease of viewing. Solid lines are fitted results based on
a model discussed in the text. Negative signals correspond to
SE/GSB in accordance with the convention in pump-probe.}
\end{figure}

To elucidate the influence of the CT states on the electronic
structure in [LuPc$_{2}$], we carried out pump-probe experiments on
[LuPc$_{2}$]$^{\bullet}$. Fig.~\ref{fig:Fig05} shows pump-probe
spectra at selected delays together with the fit based on a kinetic
model including the upper (optically allowed) exciton state, the lower
(optically forbidden) exciton state, and a hot ground state.
From the fit we conclude that the upper exciton level decays to the lower one with a 400 fs
time-constant. From this level, there is an ESA transition to a
doubly excited state. This doubly excited state is shifted by $-$550
cm$^{-1}$ as compared to twice the
LE energy. The red-shifted ESA from the dark state leads to a seemingly faster
decay of the signal on the red side of the spectrum as the
population flows into the low energy exciton state. Finally, the
lower exciton level decays into a hot ground state with a 2.5 ps
time-constant, where subsequent cooling takes place on a timescale
of 15 ps.

\subsection{Electronic Structure of [LuPc$_2$]}

To be able to interpret the experimental results on [LuPc$_{2}$]
and specifically to elucidate the role of the CT states, it is
necessary to develop a model of the electronic structure enabling
simulations of the linear- and nonlinear-spectra. Such a representation is
provided by the Frenkel exciton Hamiltonian with LE and CT states.\cite{Abramavicius2010a}
This model provides a descriptive picture
of the interaction of optical excitations with CT states, which is
useful for generalization and interpretation of the results.
Furthermore, it allows us to include doubly excited states needed to
account for ESA, which is extremely important for the interpretation
of the 2D experiments.

\begin{figure}
\includegraphics[width=8.6cm]{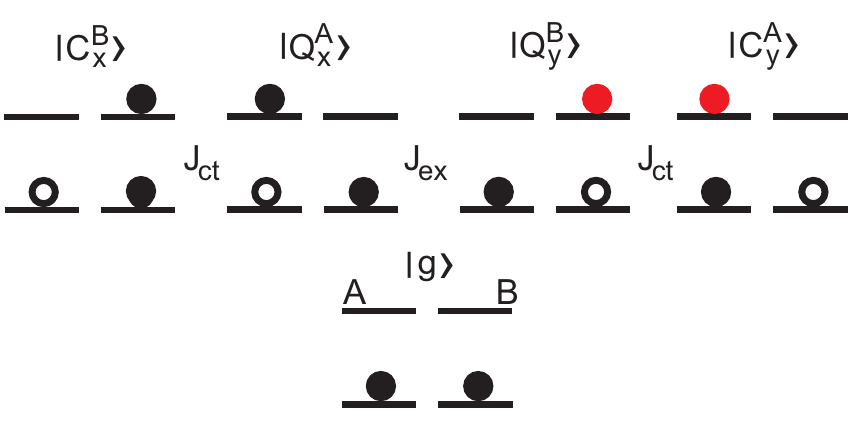}
\caption{\label{fig:Fig06} Diagrams illustrating selected singly
excited states contributing to the model. The full set of states is
obtained by permutations of A $\rightarrow$ B. Solid circles
represent electrons, hollow circles holes. Black and red color
refers to excitation of x- and y-orbitals, respectively. The
couplings between the states are indicated by $J_{ex}$ and
$J_{ct}$.}
\end{figure}

As a starting point we include the $Q_{x}$- and $Q_{y}$-excitations
on each monomer (A and B). {[}LuPc$_{2}${]} exhibits $D_{4d}$ symmetry
\cite{Koike1996}, e. g., the two rings are twisted by
45 degrees with respect to each other (Fig.~\ref{fig:Fig01}). The
transitions on the different monomers interact via resonance
coupling, and due to the overlap of the wave functions, the
excitations on the different rings are also coupled via electron
transfer into the LUMO orbitals. This allows us to construct 4
distinguishable CT states. In our terminology, $C_{x}^{A}$ is the
state where the excited electron is transferred from the excited x-
or y-orbital on monomer B to the excited x-orbital on monomer A. A
subset of the states used to construct the basis states is shown in
Fig.~\ref{fig:Fig06}. Using the configuration diagrams we can write the
Hamiltonian for the one-exciton manifold as

\begin{equation}
H_{S}=\left[\begin{array}{cccccccc}
C_{x}^{B} & 0 & J_{ct} & J_{ct} & 0 & 0 & 0 & 0\\
0 & C_{y}^{B} & -J_{ct} & J_{ct} & 0 & 0 & 0 & 0\\
J_{ct} & -J_{ct} & Q_{x}^{A} & 0 & -J_{ex} & J_{ex} & 0 & 0\\
J_{ct} & J_{ct} & 0 & Q_{y}^{A} & J_{ex} & J_{ex} & 0 & 0\\
0 & 0 & -J_{ex} & J_{ex} & Q_{y}^{B} & 0 & J_{ct} & -J_{ct}\\
0 & 0 & J_{ex} & J_{ex} & 0 & Q_{x}^{B} & J_{ct} & J_{ct}\\
0 & 0 & 0 & 0 & J_{ct} & J_{ct} & C_{y}^{A} & 0\\
0 & 0 & 0 & 0 & -J_{ct} & J_{ct} & 0 & C_{x}^{A}
\end{array}\right],\label{eq:Hs}
\end{equation}
where $J_{ex}$ is the resonance- and $J_{ct}$ is the CT-coupling.

Taking the rings as equivalent, $Q_{x}=Q_{y}$, and assuming that
$C_{x}^{A}\geq Q_{x}^{A}$, the present model recovers the previously
proposed electronic structure of [LuPc$_{2}$]$^{-}$
\cite{Ishikawa1992,Ishikawa2001,Rousseau1995}. In this model, the
band at 14200~cm$^{-1}$ has predominant charge resonance character,
while the band at 16200~cm$^{-1}$ has a dominant exciton character.
In addition, this model predicts one dark transition in the NIR
region.\cite{Ishikawa1992} This exciton model agrees very well with
our TD/DFT calculations on [LuPc$_{2}$]$^{-}$ discussed in Appendix A.
These calculations yielded three doubly degenerate electronic states with
vertical transition energies of 11682, 15848 and 18657 cm$^{-1}$, and
oscillator strengths of 0.00, 0.45 and 1.03, respectively.

Both the exciton model in the form discussed above and the direct
TD/DFT calculations predict two bright states in the relevant
spectral range, i. e. the two main absorption bands. However, none
of the models are able to explain the double peak structure of the
low energy band or the distinct dynamics of the cross-peaks below
the diagonal in the 2D spectra. To find an explanation for these
additional features found in the experiment, we investigated the
role of the counter-ion (TBA$^{+}$) present in our experiments. NMR
studies on a series of bisphthalocyanines have shown a considerable
amount of ion association in solution.\cite{Haghighi1994,Konami1989}
These experiments concluded that the counter-ion is situated on top of
one of the Pc macrocycles as indicated in Fig.~\ref{fig:Fig01}.
This is similar to the arrangement in
the crystal structure \cite{Koike1996}, which we took as a starting
point for the calculations of the optimized ground state geometries
and vertical transition energies of the
[LuPc$_{2}$]$^{-}$TBA$^{+}$ complex. The calculations show that
the interaction with the counter-ion leads to a splitting of the
doubly degenerate states found for [LuPc$_{2}$]$^{-}$, giving
rise to 6 transitions in the relevant spectral range. The energies,
oscillator strengths and contributing orbitals for
[LuPc$_{2}$]$^{-}$TBA$^{+}$ as derived from quantum chemistry
are shown in Table~\ref{tab:Tab1}.

\begin{table*}
\begin{tabular}{|c|c|c|c|}
\hline
 & $v$ (cm$^{-1}$) & \emph{f} & Orbitals\tabularnewline
\hline \hline $q_{1}$ & 10384 & 0.001 & 51\%
HOMO$\rightarrow$[LUMO$+$1], 48\%
HOMO$\rightarrow$LUMO\tabularnewline \hline $q_{2}$ & 10449 &
0.003 & 51\% HOMO$\rightarrow$LUMO, 48\%
HOMO$\rightarrow$[LUMO$+$1] \tabularnewline \hline $q_{3}$ &
15625 & 0.150 & 86\%
HOMO$\rightarrow$[LUMO$+$2]\tabularnewline \hline $q_{4}$ &
15823 & 0.156 & 85\%
HOMO$\rightarrow$[LUMO$+$3]\tabularnewline \hline $q_{5}$ &
17986 & 0.504 & 85\%
[HOMO$-$1]$\rightarrow$LUMO\tabularnewline \hline $q_{6}$ &
18116 & 0.523 & 85\%
[HOMO$-$1]$\rightarrow$[LUMO$+$1{]}\tabularnewline \hline
\end{tabular}
\caption{\label{tab:Tab1}Transition energies and oscillator
strengths for the first 6 electronic states in
[LuPc$_{2}$]$^{-}$TBA$^{+}$ at the BHLYP/SV(P) level of theory.}
\end{table*}

\begin{figure}
\includegraphics[width=8.6cm]{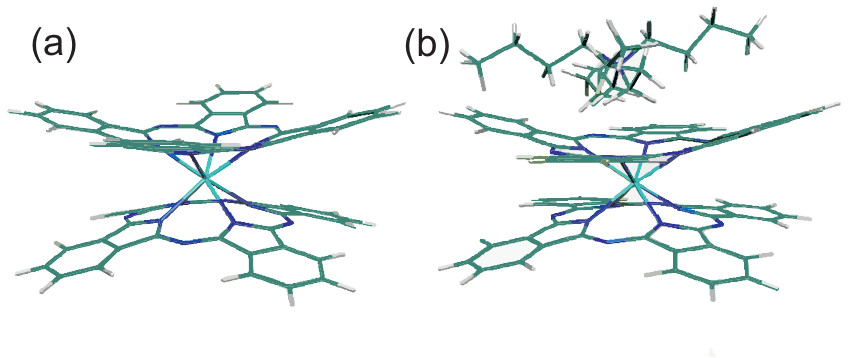}
\caption{\label{fig:Fig07}Optimized ground state geometries at the BHLYP/SV(P)
level of theory. (a) [LuPc$_{2}$]$^{-}$. (b) [LuPc$_{2}$]$^{-}$TBA$^{+}$. }
\end{figure}

The association with the counter-ion leads to significant changes in
the electronic structure of the complex and gives rise to additional
allowed electronic transitions. To understand how these changes can
be interpreted in terms of the energies and couplings between the
states in the exciton model, Fig.~\ref{fig:Fig07} shows the optimized
ground state geometry for [LuPc$_{2}$]$^{-}$ and
[LuPc$_{2}$]$^{-}$TBA$^{+}$. For [LuPc$_{2}$]$^{-}$, the
di-anionic rings (Pc$^{2-}$) display a symmetric off-planarity
distortion due to electrostatic interaction with the central metal
ion. In the presence of the counter-ion, the off-planarity distortion is decreased
significantly along the axis in contact with the alkyl chains of the
TBA$^{+}$ molecule. From an exciton model point of view, this
implies that the excitation along this axis, here $Q_{x}$, on the
ring closest to the counter-ion should be lowered in energy.
Similarly, the state corresponding to CT into the same orbital
should decrease in energy as well.

\begin{figure}
\includegraphics[width=8.6cm]{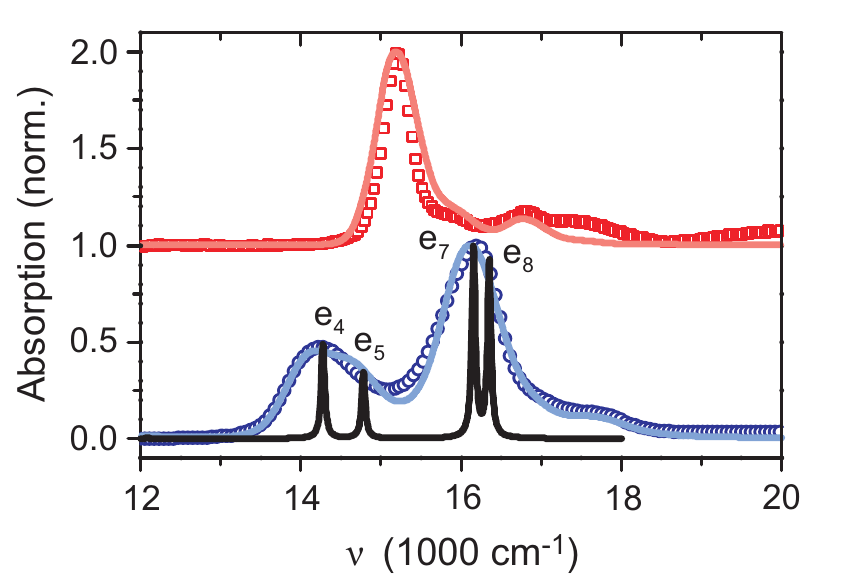}
\caption{\label{fig:Fig08}Linear absorption spectra of
[LuPc$_{2}$]$^{-}$TBA$^{+}$(blue circles) and
[LuPc$_{2}$]$^{\bullet}$ (red squares). The solid curves show
the simulations based on the model discussed in the text. For
[LuPc$_{2}$]$^{-}$TBA$^{+}$, the stick-spectrum using a single
realization of disorder is also shown. The four peaks correspond to
the transitions to the states $e_{4}$ ($14273$ cm$^{-1}$) $e_{5}$
(14778 cm$^{-1}$), $e_{7}$ (16153 cm$^{-1}$), and $e_{8}$ (16349
cm$^{-1}$) in the exciton model. }
\end{figure}

The simulations of the linear absorption spectra of
[LuPc$_{2}$]$^{\bullet}$ and [LuPc$_{2}$]$^{-}$TBA$^{+}$ based on Eq.~\ref{eq:Hs} are
shown in Fig.~\ref{fig:Fig08}. For [LuPc$_{2}$]$^{-}$TBA$^{+}$ we
also show the stick-spectrum obtained after direct diagonalization
of $H_{s}$. Denoting the ring closest to the (positive) counter-ion A,
we find that lowering the energy of $C_{x}^{A}$
(electron transfer towards the counter-ion) is able to
mimic the general features of the electronic structure found in the
quantum chemical calculations. For Fig.~\ref{fig:Fig08} we used
$Q_{x}^{A}=Q_{y}^{A}=Q_{x}^{B}=Q_{y}^{B}=$ 13500 cm$^{-1}$,
$C_{x}^{B}=C_{y}^{B}=C_{y}^{A}=$ 15150 cm$^{-1}$, $C_{x}^{A}=$ 11600
cm$^{-1}$, $J_{ex}=$ 1395 cm$^{-1}$, and $J_{ct}=$ 725.5 cm$^{-1}$.
Based on the quantum chemical calculations, we expect that the
interaction with the counter-ion should lead to a decrease of the
energy of $Q_{x}^{A}$ as well. However, changing the energy of $Q_{x}^{A}$
by $-$500 cm$^{-1}$ only leads to minor changes in the linear
absorption spectrum and we therefore kept all $Q$-transitions
degenerate for the remainder of the paper. To avoid confusion with
the states obtained from quantum chemistry, we will denote the
states of the exciton model with $e_{n}$. States $e_{1},$$e_{2}$,
and $e_{3}$ are located in the NIR and have very low or zero
transition dipole moments. States $e_{4}$ and $e_{5}$ are located in
the low energy band seen in linear absorption and are separated by
$\sim$500 cm$^{-1}$ (Fig.~\ref{fig:Fig08}). This double peak structure in
the low energy bright band is essential to reproduce the
observations in the 2D spectrum. State $e_{6}$ is dark and is
located in between the two bands, while states $e_{7}$ and $e_{8}$
are the origin of the upper band in linear absorption.

With the large number of states and couplings, a fit to the linear
absorption spectrum is by no means unique. An important test of the
validity of the model and the parameters is to apply the model
derived for the [LuPc$_{2}$]$^{-}$TBA$^{+}$ to
[LuPc$_{2}$]$^{\bullet}$. The radical itself is non-polar and
can therefore not stabilize the CT states. We modeled the spectrum
of [LuPc$_{2}$]$^{\bullet}$ by raising the energies of the CT
states to a spectral position where their influence on the $Q$
transitions is negligible. In addition, we used that the
$Q_{x}/Q_{y}$-transitions are degenerate in the absence of the
counter-ion. The successful prediction of the transition frequency,
width of the main peak, and vibrational sideband for
[LuPc$_{2}$]$^{\bullet}$ evident from Fig.~\ref{fig:Fig08} shows that
we have correctly estimated the spectral densities, resonance
coupling and $Q$-transition energies in [LuPc$_2$].

The quantum chemical calculations and the exciton model represent
two different descriptions of the electronic structure of
[LuPc$_{2}$]. The quantum chemical calculations adopt a
supra-molecular perspective and include resonance- and CT-couplings
between the two rings implicitly. The exciton model is based on
distinguishable molecular excitations with free parameters (energies
and couplings) and a minimum number of interactions. Using the language of the
exciton model, the quantum chemical calculations include more states
and couplings and will provide a more realistic molecular view. For
instance, the quantum chemical calculations account for hole
transfer and polarization of [LuPc$_{2}$]$^{-}$TBA$^{+}$ in the
ground state not included in the exciton model.
However, the quantum chemical calculations do not provide any
insight into the dynamics of the system or provide any information
about doubly excited states, and therefore the
combination of both models is needed for the interpretation of the
experimental results. The states of the two models cannot be linked
one-to-one, but a comparison of the excited state manifolds \cite{SuppMaterial}
suggests that one can well connect the dynamics of states $e_{4}$,
$e_{5}$ to $q_{3}$, $q_{4}$ and $e_{7}$,$e_{8}$ to $q_{5}$ and
$q_{6}$.

\subsection{Excited State Dynamics in [LuPc$_{2}$]$^{-}$TBA$^{+}$}

The preceding paragraph outlined a realistic description of the electronic structure of
[LuPc$_{2}$]$^{-}$TBA$^{+}$. To interpret the observed excited state dynamics we carried
out simulations of the 2D spectra based on the CT-Hamiltonian (Eq.~\ref{eq:Hs}) and
time-dependent Redfield theory in the Markov approximation (see Appendix B).
The simulated 2D spectra and kinetics are shown in the middle and bottom row in
Fig.~\ref{fig:Fig02} and Fig.~\ref{fig:Fig03}, respectively.

The simulations are in very good agreement with the experimental results,
and reproduce the shape and evolution of the 2D spectra.
At $t_{2}=$ 0, there is a clear asymmetry, where the
cross-peak below the diagonal is significantly weaker than the one above
the diagonal. It follows from the comparison of the total simulated signal
with the GSB+SE contribution (Fig.~\ref{fig:Fig02}) that the missing
cross-peak is due to overlap of ESA from population in the upper band.
The simulations also reproduce the double peak structure of the low
energy band, and we can assign this feature to the eigenstates $e_{4}$
and $e_{5}$ of the Hamiltonian $H_{s}$. The splitting between dp1 and
dp2 is close to the frequency of one of the vibrational modes found in
the quantum chemical calculations (725 cm$^{-1}$) \cite{SuppMaterial}.
If dp2 was a vibrational sideband to dp1, this mode would need to have
a Huang-Rhys factor on the order of 1 to match the observed amplitude.
This would give rise to a progression of peaks not seen in the experiments.
Furthermore, within the Condon approximation, all vibrational transitions
involving the same mode would have the same direction of the transition
dipole moment and we would thus expect to see clear cross-peaks between them. \cite{Christensson2011}
The lack of a cross-peak between dp1 and dp2 is readily reproduced in our model
(Fig.~\ref{fig:Fig02}). This can can be understood from the perpendicular orientations of the transition dipole moments of e4 and e5, which suppresses cross-peaks by a factor of three in an experiment
with all parallel polarizations.\cite{Hochstrasser2001}

\begin{figure}
\includegraphics[width=8.6cm]{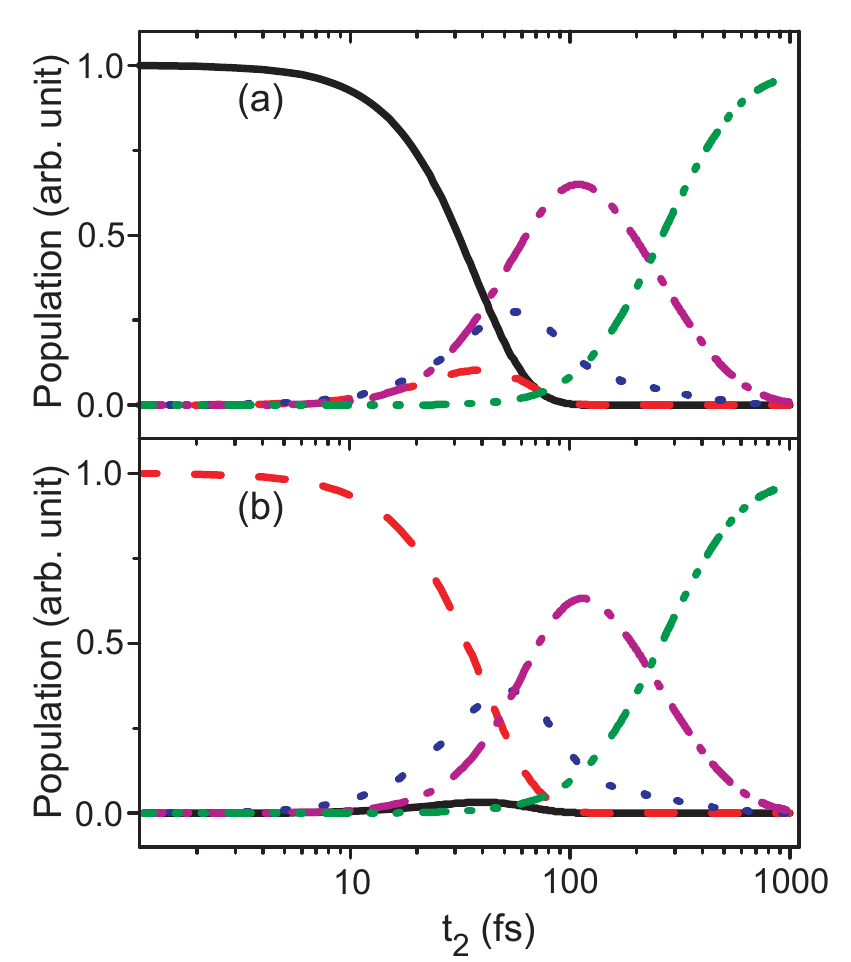}
\caption{\label{fig:Fig09}Dynamics of $e_{8}$ (black solid), $e_{7}$ (red
dash), $e_{5}$ (blue dot), $e_{4}$ (magenta dash-dot), and $e_{1}$
(olive dash-dot-dot) after initial population of $e_{8}$ (a) and
$e_{7}$ (b). }
\end{figure}

\begin{figure}
\includegraphics[width=8.6cm]{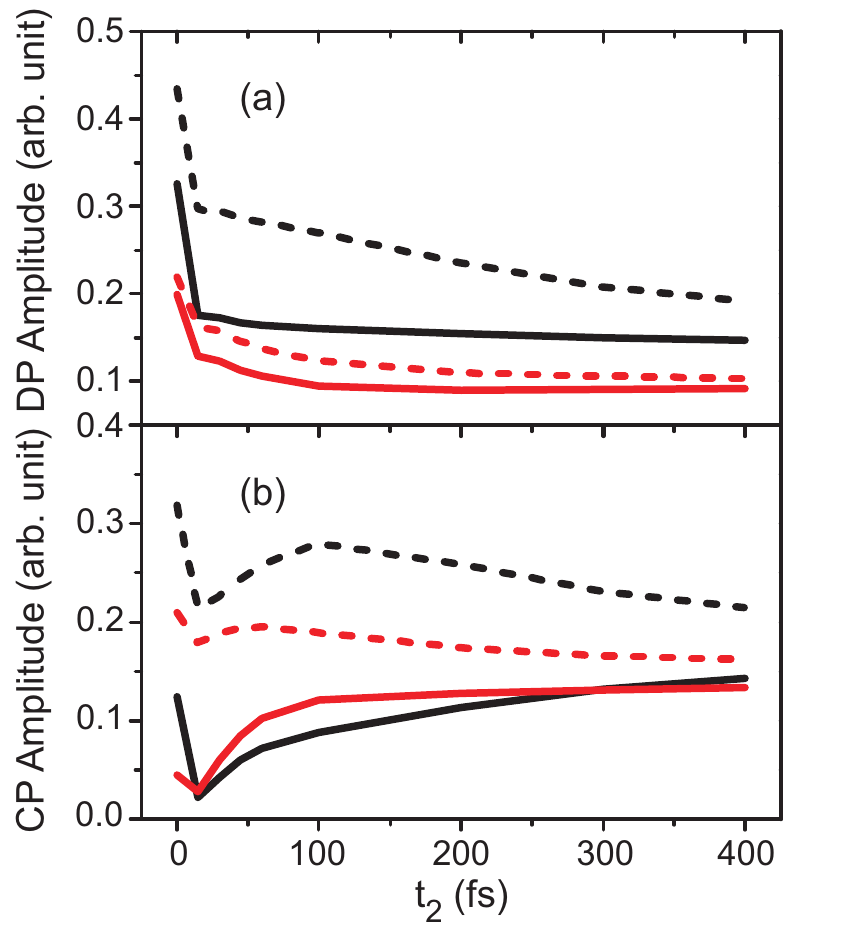}
\caption{\label{fig:Fig10}Kinetics of the simulated 2D spectra with (solid)
and without (dashed) ESA contribution. (a) dp1 (black) and dp2 (red).
(b) cp1 (black) and cp2 (red). }
\end{figure}

After excitation, coupling to the bath drives relaxation between
energy levels in the one-exciton manifold leading to dynamics in the
2D spectrum. Fig.~\ref{fig:Fig09} shows the simulated evolution of the
populations after excitation of state $e_{7}$ and $e_{8}$
corresponding to the band at 16200 cm$^{-1}$. The initially prepared
states decay and populate state $e_{5}$ within $30$ fs while the
population of state $e_{4}$ rises somewhat slower.
This is particularly obvious for initial excitation of state $e_{7}$
(Fig.~\ref{fig:Fig09}(b)), where a clear separation of the rise of
the population of $e_{4}$ and $e_{5}$ can be observed. By inspecting
the relaxation rates, we can conclude that relaxation from
$e_{8} \rightarrow e_{5}$ and $e_{7} \rightarrow e_{5}$ (Redfield rates\cite{SuppMaterial} of 18 and 11 fs)
is about 3 times faster than relaxation to state $e_{4}$ (64 and 75 fs).
This indicates that the faster rise of cp2 can be traced to faster
population relaxation into state $e_{5}$.
To investigate if this relaxation path also leads to clear signatures
in the 2D spectra, we show in Fig.~\ref{fig:Fig10} the simulated
kinetics with and without the ESA contribution. Inspecting the kinetics,
we find that the cross-peaks indeed show a fast rise reflecting SE
after population transfer from the upper band, and cp2 rises somewhat
faster than cp1. However, the SE contribution in cp2 does not acquire
significant amplitude due to fast relaxation from state
$e_{5} \rightarrow e_{4}$. Although the ESA contribution strongly
reshapes the kinetics in the 2D spectra, we can conclude that the dominating relaxation
pathway from the upper band involves relaxation via
state $e_{5}$ further on to $e_{4}$.

Turning to the diagonal peaks, we find that the SE signal decays on
a 250 fs timescale. Interestingly, this decay is almost absent when
the kinetics of the total signal is evaluated (Fig.~\ref{fig:Fig10}(b)).
This implies that the ESA contribution has a similar magnitude as
the SE contribution, and that they both decay as the population
flows out of $e_{4}$ and $e_{5}$. This matches the observations of
the experiments, where the kinetics of the low energy diagonal peak
show little dynamics even though it is clear from the pump-probe
measurements that the dark states in the bottom of the manifold of
excited states get populated on a 100~fs timescale.

The changes in the shape of the cross-peak below the diagonal during
relaxation from the upper to the lower band could provide valuable
information on the relaxation pathways. The
experimental 2D spectrum at $t_{2}=$ 45 fs (Fig.~\ref{fig:Fig02}) shows
that cp2 has its maximum at lower $v_{1}$ as compared to cp1. The
simplest interpretation of this effect is that $e_{5}$ gets
populated from $e_{7}$ while $e_{4}$ gets populated from $e_{8}$.
The kinetics shown in Fig.~\ref{fig:Fig09} shows that $e_{7}$
preferentially populates $e_{4}$ and thus gives some support for
such a conclusion. However, inspecting the simulated spectra in
Fig.~\ref{fig:Fig02}, we can conclude that the inclusion of ESA is
essential to reproduce the temporal evolution of the shape of the
cross-peaks. The negative ellipticity of the cross-peaks (an
elongation along the anti-diagonal) remains after relaxation, but is
only reproduced when the ESA contribution is included in the
simulations. The shape of the cross-peaks is thus not related to
anti-correlated diagonal disorder \cite{PisliakovFleming2006} (site
energies) or off-diagonal disorder (coupling disorder), but result
from the overlap of the different signal contributions.

\section{Discussion}

\subsection{Role of Charge Transfer States in [LuPc$_{2}$]}

The ability to alter the CT state energies in [LuPc$_{2}$]
electrochemically gives a good opportunity to study the influence of
CT states on the electronic structure and dynamics in a coupled
system. In this respect, it is instructive to compare
[LuPc$_{2}${]}$^{-}$ to [LuPc$_{2}$]$^{\bullet}$ and for a
moment neglect the influence of the counter-ion.
It is already clear from Fig.~\ref{fig:Fig01} that the presence of the
CT states in [LuPc$_{2}${]}$^{-}$ dramatically
changes the excited state structure giving rise to new transitions and a
redistribution of the oscillator strength.
Comparing the evolution of the upper state(s) in [LuPc$_{2}$]$^{\bullet}$ and
[LuPc$_{2}$]$^{-}$ illustrates that there is a strong impact on
the dynamics as well: the lifetime decreases from 400 to 30~fs in
the presence of the CT states. The results summarized in
Fig.~\ref{fig:Fig08} show that our model can interpolate between
[LuPc$_{2}$]$^{-}$ and [LuPc$_{2}$]$^{\bullet}$ by just
shifting the energy of the CT states, and this allows us to pinpoint the mechanism
for the slower dynamics in [LuPc$_{2}$]$^{\bullet}$. Our model
shows that the energetic splitting between the upper and lower band
is significantly larger in [LuPc$_{2}$]$^{\bullet}$ ($\sim2800$
cm$^{-1}$) as compared to [LuPc$_{2}$]$^{-}$ ($\sim2000$
cm$^{-1}$). We find that the fast rates in [LuPc$_{2}$]$^{-}$
can only be reproduced when we include vibrational (molecular) modes in the
bath. From this we conclude that the 1600 cm$^{-1}$ mode responsible
for the vibrational sideband in [LuPc$_{2}$] plays a central
role in the relaxation processes. To fully explain the rates in
the experiments we need to account for the influence of the
counter-ion. The association with the counter-ion splits the
degenerate energy levels and opens up additional relaxation pathways
in the appropriate energy range ($\sim1600$~cm$^{-1}$). Turning to
[LuPc$_{2}$]$^{\bullet}$, we find that the energy gap is too
large for relaxation via the high frequency vibrational mode.
Relaxation must thus proceed via the (weak) high energy wing of
the Brownian oscillator spectral density used to describe the
overdamped bath modes. The difference in the relaxation rates
between [LuPc$_{2}$]$^{-}$ and [LuPc$_{2}$]$^{\bullet}$ is
thus not related to the enhanced coupling to the bath of the CT
states, but determined by the differences in the energy level
structure of the two systems. We note that the speedup of relaxation
in the presence of CT states depends on the details of the
relaxation mechanism and the spectral densities. In the presence of
high frequency vibrational modes, the larger system-bath coupling of
the CT states may not necessarily lead to a significant increase in
the relaxation rates as was shown recently in a simulation study of
the photosystem II reaction center.\cite{Abramavicius2010a}

From the geometrical structure and linear optical properties,
[LuPc$_{2}$]$^{-}$ seems to be a highly suitable candidate for
investigations of fundamental processes like electronic coherence
dynamics, dephasing and population relaxation. Such studies would be
valuable for a better understanding of the more complicated dynamics
observed in protein pigment complexes\cite{WomickMoran2009,ZigmantasFleming2006,TurnerScholes2011}. However, the 2D spectra of
[LuPc$_{2}$]$^{-}$TBA$^{+}$ show distinct differences to the
dimer models in the literature
\cite{KjellbergPulleritsI2006,PisliakovFleming2006,EgorovaDomckeII2007,Voronine2006,Abramavicius2010b,Yuen-Zhou2011,ChoFlemingII2005,Chen2010}.
These differences cannot be explained by a rapid damping of the
electronic coherences due to stronger system-bath coupling of the CT
states. Rather, our analysis shows that including
vibrational modes as well as employing a proper model for the two-exciton
manifold is needed in order to understand the spectra and dynamics.
The properties of the states in the two-exciton manifold responsible
for ESA cannot be deduced from available quantum chemical
calculation with the needed precision, and therefore we need to
resort to the exciton model. Perhaps the most prominent observation
in the present experiments is the absence of a cross-peak below the
diagonal at short population times (or lack of SE in pump-probe).
While this requires a special structure of the two-exciton manifold,
the same effect can show up in any coupled system. In fact, such
cancelation happens readily in the exciton model when the two LEs
have unequal transition dipole moments. The absence of the
cross-peak should thus not be seen as a direct effect of the CT
states, but illustrates the importance of the two-exciton manifold
and ESA for a proper interpretation of the spectra and dynamics.
This might seem surprising because the 2D spectra are dominated by
positive signal contributions (Fig.~\ref{fig:Fig02}). However, ESA
reshapes the peaks and strongly affects the observed kinetics
(Fig.~\ref{fig:Fig10}). We note that our simple model for the two-exciton
band can account for the cancelation of the cross-peak below the
diagonal as well as the absence of decay dynamics in the low energy
band. To reach this agreement we need to shift the two-exciton band
by $-$1000 cm$^{-1}$ from the values estimated from the Frenkel
exciton model. A similar redshift of the two-exciton manifold
can be observed in [LuPc$_{2}$]$^{\bullet}$ ($-$550 cm$^{-1}$),
indicating that electron correlation effects
\cite{Raghavachari1996} are significant in this system even in the
absence of the CT states. Inspecting the (full) pump-probe spectrum, one finds a very broad ESA covering the entire visible spectral range.\cite{SuppMaterial}
In order to reproduce the ESA spectrum in the visible spectral range, we would need to include
more (high energy) configurations to the two-exciton Hamiltonian.
The interaction of these high energy configurations with the ones
included in our model could further contribute to the red-shift of
the ESA observed in the experiments.

The current experiments and analysis highlight the
importance of the two-exciton manifold for the interpretation of
time-resolved spectroscopy of processes which take place exclusively
in the one-exciton manifold. In one-dimensional techniques, like
pump-probe or transient grating, the presence of ESA may easily be
overlooked leading to erroneous interpretation of the dynamics. In this respect, the
combination of single- and double-quantum 2D spectroscopy is most useful to
elucidate the role of ESA and the associated signal contributions.\cite{NemethChristensson2010,Christensson2010}

The pronounced influence of ESA in [LuPc$_{2}$] makes in-depth analysis
of the 2D spectra without numerical modeling difficult. Such simulations
require a simple Hamiltonian mimicking the electronic structure of the
system. There are many models which could fit the linear spectrum and
give rise to a suppressed cross-peak below the diagonal which grows
as $t_2$ increases, e. g., the simple exciton model with
unequal transition dipole moments of the two LEs or the isolated
[LuPc$_{2}$]$^-$ model discussed above. However, none of these models
are able to reproduce the dynamics in the cross-peak below the diagonal,
or the double peak structure of the low energy band. By linking the
exciton model to quantum chemical calculations, it is clear that
precisely these features are signatures of special
interactions, and that they provide important information on the
dynamics in the system.

\subsection{Excited State Charge Transfer in [LuPc$_{2}$]$^{-}$TBA$^{+}$}

The association of [LuPc$_{2}$]$^{-}$ with the TBA$^+$ counter-ion provides
a natural explanation for the observation of the double peak structure in the
low energy band in the 2D spectra (Fig.~\ref{fig:Fig02} and Fig.~\ref{fig:Fig08}).
The formation of such ion-pairs has strong impact on the character of the excited states.
The (positive) charge of the counter-ion serves to stabilize states representing CT towards the counter-ion. This gives rise to an asymmetry in the energies of the CT states,
and as a result, the electronic eigenstates in [LuPc$_{2}$]$^{-}$TBA$^+$ will
reflect varying degrees of charge separation.
Relaxation between the different energy levels thus
represents a net redistribution of electron and hole densities between the
two rings, which can be followed by the evolution of the cross-peaks in
the 2D spectra. By combining the analysis of the cross-peak dynamics within
the exciton model with the results of the quantum chemical calculations, it becomes possible to obtain a molecular view of the CT dynamics.

\begin{figure}
\includegraphics[width=8.6cm]{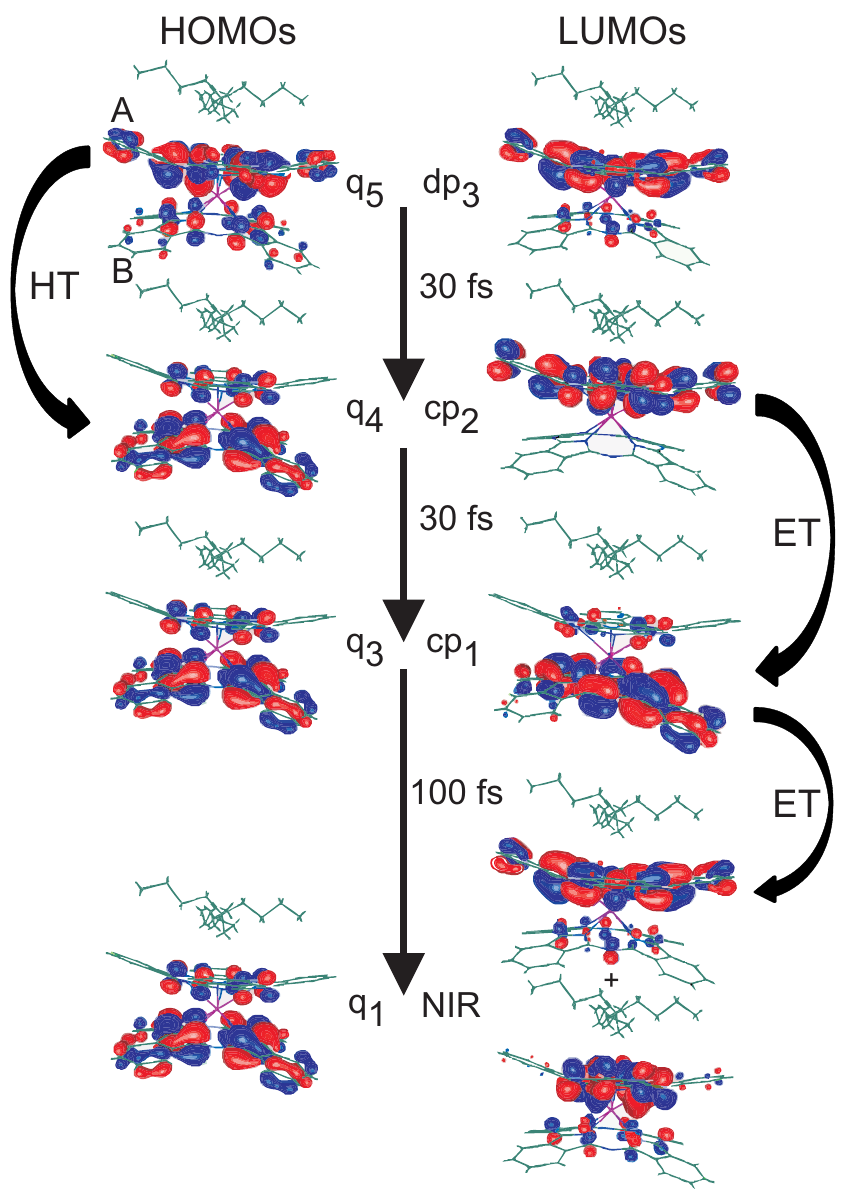}
\caption{\label{fig:Fig11}Dominant orbitals involved in selected
transitions for [LuPc$_{2}$]$^{-}$TBA$^{+}$ using
TD-BHLYP/SV(P). The legend to the left refers to the number of the
state in Table~\ref{tab:Tab1}, and the legend to the right to the features
assigned in Fig.~\ref{fig:Fig02}. The timescales are obtained from the
analysis of the kinetics of the 2D spectra. The curved arrows
indicate the different electron- (ET) and hole-transfer (HT)
processes taking place after excitation of the upper band in
[LuPc$_{2}$]$^{-}$TBA$^{+}$.}
\end{figure}

Fig.~\ref{fig:Fig11} shows the orbitals for the different bright transitions
with labels indicating the states and the corresponding features in the spectrum in
Fig.~\ref{fig:Fig02}. In addition, curved arrows are used to indicate the
CT processes connecting the different states. An excitation of the
high energy band (dp3) populates an excited state with most of the
charge density located on the side of the counter-ion (ring A). The
system first relaxes via hole transfer (CT in the HOMO orbital) to
ring B in 30 fs. This process is revealed via the faster rise of cp2
in the 2D spectrum. Further relaxation requires electron transfer
(CT in the LUMO orbital) to ring B, and we find that this process
also takes place on a 30 fs timescale. The system then goes to a
stable (on ps timescale) charge separated state with a transition in
the NIR. For this state, the excited state charge density is located
on the ring closest to the counter-ion. Our pump-probe measurements
reveal that population of the dark states in the NIR takes only
about 100 fs. The combination of CT states and asymmetric
interaction with the counter-ion thus opens up a relaxation pathway
which very rapidly localizes the charge on one side of the complex.
At first glance, one may assume that the counter-ion lowers the energy
of both $C_{x}^{A}$ and $C_{y}^{A}$ and stabilizes the charge in both
orbitals on the side of the counter-ion. However, it is clear from both
the quantum chemical calculations as well as the exciton model that the
molecular details of the interaction between [LuPc$_{2}$]$^{-}$ and the
counter-ion need to be considered. The counter-ion induces different
distortions of the molecular structure along the x- and y-coordinates,
and this also affects the energies of the CT states representing CT
into the corresponding orbitals on the macrocycle closest of the counter-ion. This selectivity
is clearly manifested in the orbitals shown in Fig.~\ref{fig:Fig11}, and
shows that a detailed understanding of the ultrafast dynamics requires
careful consideration of the interactions in the ion-pair.

[LuPc$_{2}$]$^{-}$TBA$^{+}$ has many similarities to the so called special
pair in the bacterial reaction center. The special pair is a dimer
of bacteriochlorophylls coupled to CT states with higher transition energy
than the fundamental excitations.\cite{Renger2004}
However, because of different orientation of the transition dipoles,
the lowest excited state (often named P$^{-}$)
carries most of the oscillator strength. Charge separation in the
special pair takes place after direct excitation or energy transfer to P$^{-}$.
Numerous works have concluded that P$^{-}$ corresponds to a state with
most of the charge density located on the M-side of the special
pair. This localization of the charge density has been explained by
specific interactions, such as hydrogen bonds
\cite{Moore1999,Parson1987}, asymmetric interaction with the protein
\cite{Gunner1996,Steffen1994}, or a special arrangement and
interaction of the reaction center pigments
\cite{Kolbasov2000,Parson1987,Scherer1989,Ren2009}. In the language
of the exciton model, all these mechanisms serve to lower the CT
states on one side of the complex in a similar way as the
counter-ion in our case. Our results show that the presence of an
asymmetric interaction facilitates relaxation to the bottom of the
band, where the charge is localized on one side of the complex. In
the special pair, the relaxation time between the upper and lower exciton
band has been measured to be 65 fs.\cite{ArnettSchererI1999} The
localization of the charge density on the M side of the special pair
is thought to be important for the directionality of electron
transfer in the bacterial reaction center, since the M side has a
better overlap with the accessory bacteriochlorophyll on the L
branch \cite{Ren2009}. Our experiments
show that this type of charge localization can be mimicked by our
model system, and followed in real time with 2D spectroscopy.
The combination of 2D spectroscopy and theoretical analysis including
density matrix propagation and quantum chemical calculations thus offers
new possibilities to disentangle the complicated CT dynamics observed in
the photosystem II reaction center \cite{MyersOgilvie2010, LewisOgilvie2012}.

\section{Conclusions}

In this work we have investigated the electronic structure and excited state dynamics
in a bis-phthalocyanine dimer ([LuPc$_{2}$]$^{-}$) where the resonance- and CT-coupling
are of the same magnitude. To selectively investigate the effects from the resonance coupling,
we have exploited the possibility to "turn off" the CT coupling by oxidation
of the complex ([LuPc$_{2}$]$^{\bullet}$).
Based on linear absorption,
2D spectroscopy and pump-probe measurements, we have shown that the mixing of
exciton and CT states in [LuPc$_{2}$]$^{-}$ leads to significant changes of
the electronic structure and a tenfold speed-up of the excited state dynamics as
compared to the excitonic [LuPc$_{2}$]$^{\bullet}$. The speed-up of the
dynamics in [LuPc$_{2}$]$^{-}$ can be qualitatively understood from the
changes in the energy level spacing due to the presence of CT states in combination
with quasi-resonances between the energy gaps and high frequency vibrational modes.
However, a detailed analysis of the 2D spectra has shown that the interaction of
[LuPc$_{2}$]$^{-}$ with its TBA$^{+}$ counter-ion needs to be considered to
account for all peaks and their evolution. Using quantum chemical calculations,
we have demonstrated that the specific details of the interaction in the ion-pair
determines the electronic structure of [LuPc$_{2}$]$^{-}$TBA$^+$.
The interaction with the counter-ion stabilizes CT states corresponding to CT into specific orbitals on the macrocycle closest to the counter-ion. The subsequent imbalance in energy of the
CT states results in an electronic structure where the different excited states
represent varying degrees of charge separation. Relaxation in the excited state
of the complex thus proceeds via electron- and hole-transfer processes,
which give rise  to distinct cross-peak dynamics in the 2D spectra.
The analysis of the cross-peak dynamics, together with quantum chemical calculations,
demonstrate the ability of 2D electronic spectroscopy to provide a detailed molecular view on these
transient CT processes.

\begin{acknowledgments}
This work was supported by the Austrian
Science Foundation (FWF), projects P223311 and F016-18 (SFB ADLIS),
\"Osterreichischer Austauschdienst (OeAD, WTZ CZ07/2011), and the
Deutsche Forschungsgemeinschaft through the DFG-Cluster of
Excellence Munich-Centre for Advanced Photonics. V.L. thanks for the
opportunity to use the computer facilities at the University of
Vienna (Schr\"odinger Cluster) and at the Institute of
Theoretical Chemistry, University of Vienna. T.M. acknowledges
support by the Czech Science Foundation (GACR) through grant No.
205/10/0989 and by the Ministry of Education, Youth, and Sports of
the Czech Republic through grant KONTAKT ME899 and the research plan
No. MSM0021620835. N.C. acknowledges support from the Wenner-Gren
foundation.
\end{acknowledgments}

\vspace{1cm}

\appendix{\label{AppendixA}\textbf{Appendix A. Quantum Chemical Calculations}}

The BHLYP \cite{Becke1993a} functional was applied in Density
Functional Theory (DFT) and time-dependent (TD-)DFT
\cite{Bauernschmitt1997} calculations of optimal electronic ground
state and lowest singlet excited-state geometries. This functional
combines Becke\textasciiacute{}s half-and-half exchange functional
with the LYP correlation functional proposed by Lee, Yang and Parr.
Among the tested functionals (PBE
\cite{Perdew1996}, B3LYP \cite{Becke1993b}, and BHLYP \cite{Becke1993a}), the overall best
results were obtained using BHLYP, which can be explained by the
larger amount of exact exchange \cite{Dreuw2004}. The optical
transitions were calculated using (TD)-DFT on the basis of the
optimized geometries. We employed the split valence basis sets
def-SV(P) (for N, C, H and O atoms) and def-ecp for lutetium
\cite{Dunning1977} from the TURBOMOLE library. Herein, valence
orbitals were calculated in a double-zeta basis and augmented with
higher angular momentum polarization functions to account for the
nonplanarity of the macrocycles. All quantum chemical calculations
were performed with the Turbomole 5.7 package \cite{Ahlrichs1989}.

\vspace{1cm}

\appendix{\label{AppendixB}\textbf{Appendix B. Simulations of Excited State Dynamics and Non-linear Spectra}}

Excited state dynamics were simulated by applying time-dependent
Redfield theory to the diagonalized Frenkel exciton Hamiltonian of
the system.\cite{MayKuhn2000} Altogether eight singly excited
electronic states were considered in the {[}LuPc$_{2}${]} dimer. In
addition to the local $Q_{x}/Q_{y}$-excitations on each Pc ring, CT
states corresponding to electron transfer from the excited state
orbital on one macrocycle to the excited state orbital on the other
ring were considered. The CT states were modeled as excited states
with zero transition dipole moment from the electronic ground state
and increased reorganization energy (with respect to the LE states).\cite{Renger2004,Mancal2006}
The sign of the resonance coupling matrix elements, $J_{ex}$, was
determined from the structure. For the coupling between CT and LE states,
$J_{ct}$, we chose the signs of the matrix elements so that the
Hamiltonian was invariant under rotation of the rings. The magnitudes
of $J_{ex}$ and $J_{ct}$ were fitted to the linear absorption spectrum.
Linear and non-linear spectra were calculated by a standard semiclassical
response function theory.\cite{Mukamel1995} For calculation of
non-linear spectra we constructed response functions corresponding
to GSB, SE, and ESA in the Markov approximation
\cite{ZhangMukamelI1998}, i.e. neglecting correlations between the
systems time evolution in different intervals of the response
functions. Orientational average was directly taken into account for
each response function.\cite{Hochstrasser2001} To describe line
shapes and kinetics of the peaks, the dissipative population and
coherence dynamics have to be calculated for the electronic states
of the molecule. Dissipation was included via an analytical form of
the time-dependent Redfield equation, invoking the secular
approximation.\cite{SuppMaterial} For the calculation of the
nonlinear spectra, the two-exciton manifold responsible for the ESA
contributions needs to be included. As a first approximation we
included the 28 states which can be constructed by simultaneous
excitation of two single exciton transitions (doubly excited
states). This is a straightforward generalization of the usual
Frenkel exciton model to the case of CT states. However, electron
correlation effects \cite{Raghavachari1996}, or inclusion of doubly
excited monomeric states \cite{Pullerits1996} can shift the energy
of the two-exciton manifold. To account for these effects, not
included in our model, we adjusted the energy of the two-exciton
band to obtain the best agreement between simulation and experiment.
A more rigorous treatment of the two-exciton
CT band would have to account for various special properties of the
combined doubly excited states, e.g., the fact that some double CT
configurations result in normal double exciton states.\cite{Abramavicius2010a}
For calculation of the dephasing in the ESA part of the signal, the depopulation rates of the doubly-excited states are required. Here
we assumed that the depopulation rates are equal in the one- and
two-exciton manifolds.

The coupling of the electronic transitions
to the bath, described by the spectral density, determines both the
line shapes and the population relaxation rates. Each transition was
coupled to 3 vibrational modes with frequencies of 160, 725, and
1600 cm$^{-1}$ and Huang-Rhys factors of $S=$ 0.3, 0.4, and 0.3,
respectively. Additionally, one over-damped Brownian
oscillator with reorganization energy of 80 cm$^{-1}$ and a decay
rate of 100~fs$^{-1}$ was used. For the CT states, we increased
the coupling to the over-damped mode by a factor of 1.3. The
vibrational modes are needed in order to reproduce the vibrational
sideband in linear absorption as well as the fast rates found for
[LuPc$_{2}$]$^{-}$. In all calculations of optical spectra, we
explicitly averaged the signals over 1000 configurations where the
transition energies were randomly sampled from a Gaussian
distribution with a FWHM of 300 cm$^{-1}$.

\newpage

%

\end{document}


\title{Supporting Information\\Ultrafast Photo-Induced Charge Transfer Unveiled by\\ Two-Dimensional
Electronic Spectroscopy}

\author{Oliver Bixner$^1$, Vladim\'{i}r Luke\v{s}$^2$, Tom\'{a}\v{s} Man\v{c}al$^3$, J\"{u}rgen~Hauer$^1$, Franz~Milota$^1$,Michael~Fischer$^4$,\\ Igor~Pugliesi$^5$, Maximilian~Bradler$^5$, Walther~Schmid$^4$,Eberhard~Riedle$^5$,
Harald~F.~Kauffmann$^{1,6}$,\\ and Niklas~Christensson$^1$}

\affiliation{$^1$Faculty of Physics, University of Vienna,
Strudlhofgasse 4, 1090 Vienna, Austria\\
$^2$Department of Chemical Physics, Slovak Technical
University, Radlinsk\'{e}ho 9, 81237 Bratislava, Slovakia\\
$^3$Institute of Physics, Faculty of Mathematics and
Physics, Charles University, Ke Karlovu 5, Prague 121 16, Czech
Republic\\
$^4$Department of Organic Chemistry, University of Vienna,
W\"{a}hringer Straße 38, 1090 Vienna, Austria\\
$^5$Lehrstuhl f\"{u}r BioMolekulare Optik,
Ludwig-Maximilians-University, Oettingenstrasse 67, 80538 Munich,
Germany\\
$^6$Ultrafast Dynamics Group, Faculty of Physics, Vienna University
of Technology, Wiedner Hauptstrasse 8 - 10, 1040 Vienna, Austria}

\maketitle
\tableofcontents{}
\newpage

\section{Synthesis}
A mixture of lutetium acetate (5 mmol), 1,2-o-dicyanobenzene, and
sodium carbonate in a stoichiometric ratio of 1 : 8 : 0.5 were
stepwise heated to 280~$^{\circ}$C and kept at constant temperature
until solidification of the melt signaled completion of the
reaction. The reaction product was purified by sublimation under
reduced pressure ($<$~1~Torr), following chromatographic separation
and recrystallization.\cite{Koike1996,Konami1989} Column
chromatography was carried out by dissolving the crude reaction
mixture in a minimum amount of acetonitrile and flushing it over
aluminium oxide (AlO$_x$ 90, neutral, activity grade I) using sodium
methanolate (MeOH / MeO$^-$Na$^+$ (0.2 \%)) as basic, polar mobile
phase. The complex was precipitated by addition tetrabutylammonium bromide (TBABr) and crystallized for 3 days at -30$^{\circ}$C. Recrystallization
from DMF yielded [LuPc$_2$]$^-$TBA$^+$ (5 \%). All reagents were
purchased from Sigma Aldrich and used as received without further
purification. Solvents were applied in absolute forms.

The complex was characterized using a number of spectroscopic
techniques listed below.

\begin{tabbing}

ESI-MS: (CH$_3$OH/CH$_3$CN) m/z: \quad\= 0.96 (t), 1.31
(m),1.57(m),3.17 (s), 8.20 (q), 8.86 (q) \kill
$^1$H-NMR: (DMSO) $\delta$ (ppm): \> 0.96 (t), 1.31 (m), 1.57 (m), 3.17 (s), 8.20 (q), 8.86 (q) \\
$^{13}$C-NMR: (DMSO) $\delta$ (ppm): \> 13.84 (CH$_3$), 19.57 (CH$_2$), 23.41 (CH$_2$), 57.89 (CH$_2$), \\
\> 121.91 (CH), 129.00 (CH), 137.15 (Q), 158.96 (Q) \\
UV/VIS: (BN) $\nu_{max}$ (cm$^{-1}$): \> 14245, 16180, 17700, 24330, 30030 \\
IR: (BN) $\nu_{max}$ (cm$^{-1}$): \> 1112, 1330, 1381, 2875, 2935, 2965 \\
Ra: (BN) $\nu_{max}$ (cm$^{-1}$): \> 1505, 3050 \\
ESI-MS: (CH$_3$OH/CH$_3$CN) m/z: \> 1199.8 (M$^+$) \\
\end{tabbing}

The radical was generated by dissolving [LuPc$_2$]$^-$TBA$^+$ in
methanol and adding an equal amount of water and 0.5 \%v/v
concentrated hydrochloric acid. After evaporation of the solvent,
the precipitate containing the neutral radical was isolated and
dissolved in toluene.

\begin{tabbing}
ESI-MS: (CH$_3$OH/CH$_3$CN) m/z \quad\= 0.96 (t), 1.31
(m),1.57(m),3.17 (s), 8.20 (q), 8.86 (q) \kill
UV/VIS: (Toluene) $\nu_{max}$ (cm$^{-1}$): \> 15190, 15950, 16830,
17630
\end{tabbing}

\newpage

\section{Phasing of the 2D Experiment}

\begin{figure}
\includegraphics[width=0.65\textwidth]{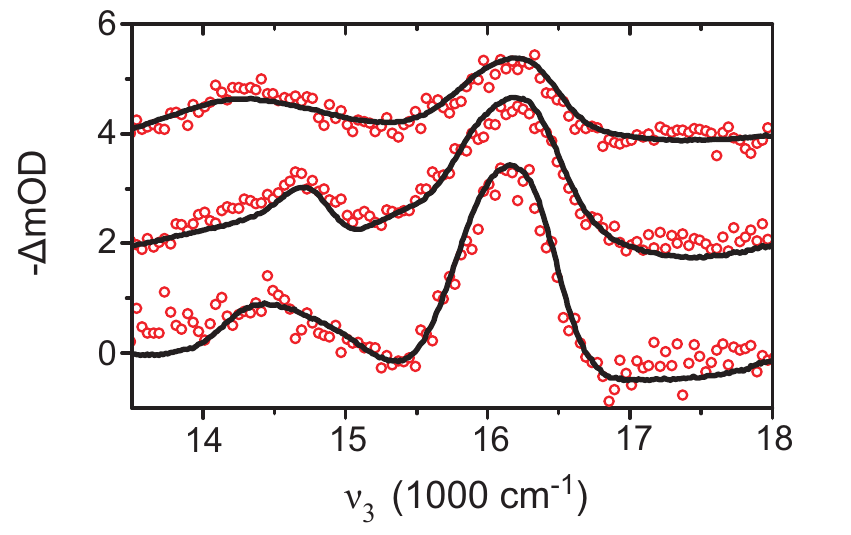}
\caption{\label{figS01} Spectrally resolved pump-probe signal (red
circles) and the projection of the real part of the 2D spectrum
(solid) for $t_2=$ 0 fs, 45 fs, and 400 fs (from bottom to top). The
spectrum for $t_2=$ 45 fs (400~fs) has been displaced by 2 (4) mOD
along the vertical axis for clarity.}
\end{figure}

To extract the real part of the signal we employ a phasing procedure
described previously.\cite{BrixnerFlemingII2004} Accurate phasing
is crucial for obtaining reliable information from the shape and
dynamics of the various peaks. We find that all spectra need to be
phased to their corresponding pump-probe spectrum to avoid errors
due to the drift of the phase between the measurements. Fig.~\ref{figS01}
shows the comparison of the pump-probe signal and the projection of
the real part of the 2D spectrum for a few selected $t_2$ times. The
phasing produces good agreement over the entire spectral range and
for all times.

\section{Quantum Chemistry}

\begin{figure*}
\includegraphics[width=0.95\textwidth]{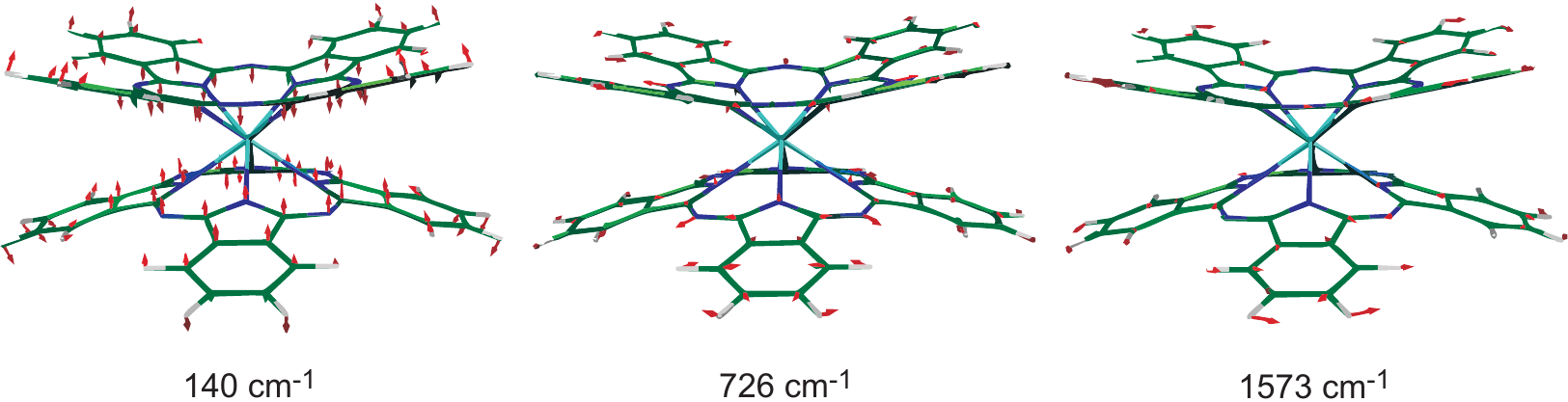}
\caption{\label{figS02} Motions of the atoms for the normal modes in Table
\ref{tabS01}.}
\end{figure*}

Normal mode analysis of [LuPc$_2$]$^-$ revealed a number of normal
modes with significant Huang-Rhys factors with respect to the
electronic state corresponding to the low energy band in linear
absorption. The most striking mode is the 140~cm$^{-1}$ (160
cm$^{-1}$ in experiments) mode involving a modulation of the
ring-to-ring distance. The motions of the most prominent normal
modes are shown in Fig. \ref{figS02} and the Huang-Rhys factors and
frequencies are summarized in Table \ref{tabS01}.

\begin{table}
  \centering
  \begin{tabular}{|c|c|}
    \hline
    $\nu$ cm$^{-1}$ & Huang-Rhys factor \\
\hline
    140 & 0.5604 \\
\hline
    726 & 0.0697 \\
\hline
    1573 & 0.31 \\
    \hline
  \end{tabular}
  \caption{Dominant ground-state vibrational frequencies and Huang-Rhys (HR)
  factors for [LuPc$_2$]$^-$ calculated with TD-DFT (BHLYP/SV(P)).}\label{tabS01}
\end{table}

\begin{figure}
\includegraphics[width=0.65\textwidth]{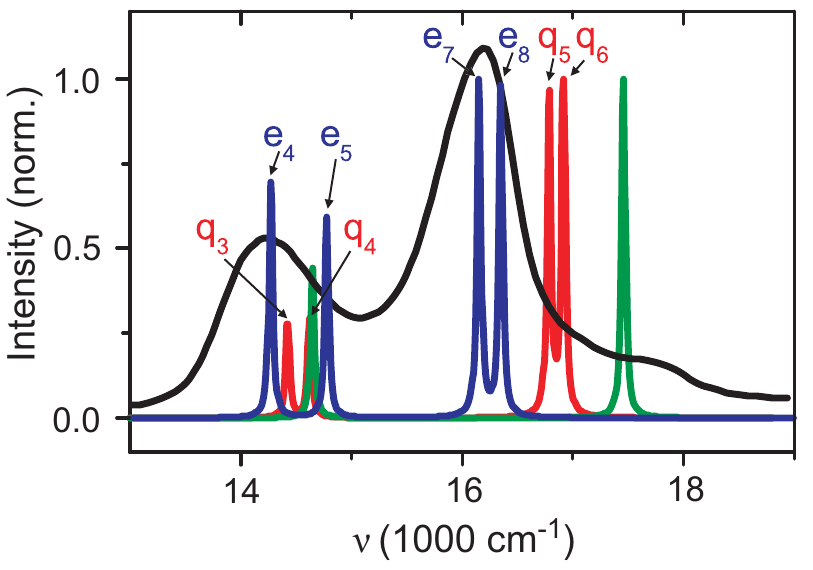}
\caption{\label{figS03} Transitions energies from the exciton model
(blue) as compared to those of quantum chemistry for the
[LuPc$_2$]$^-$TBA$^+$ (red) and [LuPc$_2$]$^-$ (green). Both quantum
chemical calculations have been shifted by $-$1200 cm$^{-1}$. The
labels show the respective states in the exciton model and quantum
chemical calculations. Also shown is the linear absorption spectrum
of [LuPc$_2$]$^-$.}
\end{figure}

Neither the exciton model nor the quantum chemical calculations are
sufficient for a complete interpretation of the experimental
results. The exciton model is needed for the interpretation of the
excited state dynamics and simulations of the 2D spectra, but
provides little insight into the molecular nature of the
transitions. The quantum chemical calculation on the other hand
provides a more realistic picture of the electronic structure of the
molecule, especially in the presence of the counter-ion.
Fig. \ref{figS03} shows a comparison of the vertical transition energy
for the first few excitations in the exciton model and the quantum
chemical calculations with and without the counter-ion. The quantum
chemical calculations have been shifted by $-$1200 cm$^{-1}$. In the
presence of a solvent, the two Pc rings are expected to planarize
due to polarization effects leading to a red-shift of the
transitions. The addition of the counter-ion leads to a splitting of
the doubly degenerate bands and a red-shift of all transitions. We
find that the splitting between the upper and lower band is somewhat
larger in the calculations as compared to the exciton model. The
similarities between the spectra of the two models suggest that
$e_4$ and $e_5$ of the exciton model can be interpreted as $q_3$ and
$q_4$ from the quantum chemical calculations (as well as $e_7$,
$e_8$ and $q_5$, $q_6$). A direct comparison of the transitions of
the two models is complicated by the fact that the exciton model
assumes that the ground state wave-function is evenly distributed
over the two monomers. From the quantum chemical calculations, we
can see that the interaction with the counter-ion leads to a
significant localization of the ground charge density on the side of
the counter-ion. Such effects are beyond our simplified model, and
therefore we use the orbitals from the quantum chemical calculations
to interpret the re-distribution of charge following relaxation.

\section{Time Dependent Redfield Rates}

Relaxation processes in the first excited band of the dimer are
described by the second order quantum master equation in the secular
approximation. The system-bath interaction Hamiltonian is written in
product form
\begin{equation}\label{eq01}
    H_{S-B} = \sum_n \Delta\Phi_nK_n
\end{equation}
where $\Phi_n$ (the energy gap operator) is the difference of the
environmental potential energy surfaces of the ground state and the
excited state $|n\rangle$. The operator $K_n$ is the projector
formed of the states $|n\rangle$, i.e. $K_n=|n\rangle\langle n|$.
Starting with the well-known Nakajima-Zwanzig identity, the
relaxation term in the equations of motion for the reduced density
matrix $\rho(t)$ takes the form of an integral \cite{MayKuhn2000}
$$
    \frac{\partial \rho}{\partial t} =
    -\sum_n\int^{t-t_0}_0d\tau\{C_{nn}(\tau)[K_n,U_S(\tau)K_n\rho(t-\tau)U^+_S(\tau)]_{-}$$
\begin{equation}\label{eq02}
    - C_{nn}(-\tau)[K_n,U_S(\tau)\rho(t-\tau)K_nU^+_S(\tau)]_{-}\}~.
\end{equation}
In the integral on the right-hand side we introduce slow envelop
approximation $U^+_S(t-\tau)\rho(t-\tau)U_S(t-\tau)\approx
U^+_S(t)\rho(t)U_S(t)$, by which the integro-differential Eq. \ref{eq02}
is converted into an ordinary differential equation with a
time-dependent tensor. In the electronic eigenstate representation
it reads
\begin{equation}\label{eq03}
    \frac{\partial \rho_{ab}(t)}{\partial
    t}|_{relax}=\sum_{cd}R_{abcd}(t)\rho_{cd}(t),
\end{equation}
where
\begin{equation}\label{eq04}
    R_{abcd}(t)=\delta_{ab}\sum_e\Gamma_{beed}(\omega_{de};t)+\delta_{bd}\sum_e\Gamma_{aeec}(\omega_{ce};t)-\Gamma_{cabd}(\omega_{db};t)-\Gamma_{dbac}(\omega_{ca};t).
\end{equation}
The individual components of the tensor read
\begin{equation}\label{eq05}
    \Gamma_{abcd}(\omega;t) = \sum_n\langle a|K_n|b\rangle\langle
    c|K_n|d\rangle \mathrm{Re}\int^{t-t_0}_0d\tau
    e^{i\omega\tau}C_{nn}(\tau)~.
\end{equation}
In addition, we neglect all terms coupling the evolution of
coherences and populations (secular approximation). In a long time
limit, i.e. if $t_0 \rightarrow \infty$, this procedure leads to the
well-known Redfield equations \cite{MayKuhn2000}. For our present
purposes, we have to take into account the transient period, when
the relaxation rates "build up", i.e. when they are time dependent.
As we can see in Eq. \ref{eq06}, the Redfield tensor is composed of
terms exhibiting the following form
\begin{equation}\label{eq06}
    \Gamma(\omega;t)=\alpha\mathrm{Re}\int^t_0d\tau e^{-i\omega
    \tau}C(\tau)~,
\end{equation}
where $\alpha$ is a constant and $\omega$ is a transition frequency
in the excitonic band (we set $t_0=$ 0). To facilitate the
calculations, we evaluate Eq. \ref{eq06} for the case of a single
overdamped Brownian oscillator with $C(t)=C_0e^{-\gamma t}$, where
$\gamma$ is the inverse of the bath correlation time. We obtain
\begin{equation}\label{eq07}
    \Gamma(\omega;t)=\alpha\frac{c_0\gamma}{\gamma^2+\omega^2}\Big{(}1-e^{-\gamma t}
    (\cos(\omega t)+\frac{\omega}{\gamma}\sin(\omega t))\Big{)}~,
\end{equation}
which leads in the long time limit to
\begin{equation}\label{eq08}
    \Gamma_{\infty}=\alpha\frac{c_0\gamma}{\gamma^2+\omega^2}~.
\end{equation}
the $\Gamma_{\infty}$ are the components of the standard Redfield
relaxation tensor. To capture the essential feature of the time
dependence of the rates, we neglect the frequency of the transition
with respect to the bath correlation time, i.e. we assume
$\cos\omega t \approx 1, \sin\omega t\approx 0$ if $e^{-\gamma
t}>0$. Neglecting the dependence of $\Gamma(\omega;t)$ on $\omega$,
the Redfield tensor component is given by
$\Gamma(\omega;t)=\Gamma_{\infty}(1-e^{-\gamma t})$. For times
longer than the bath correlation time we recover the standard (time
independent) Redfield component $\Gamma_{\infty}$. Assuming that the
correlation functions of all components decay with the same constant
$\gamma$, the time-dependent relaxation tensor reads
\begin{equation}\label{eq09}
    R(t)=R_{\infty}(1-e^{-\gamma t})~,
\end{equation}
where $R_{\infty}$ is the standard Redfield tensor. The form of the
relaxation tensor implies that time dependent rates can combine fast
relaxation (long time rates) with relatively narrow homogenous
linewidths (initial rates). Furthermore, the ansatz in Eq. \ref{eq09}
greatly simplifies the calculation of the evolution superoperators
needed to propagate the equation of motion. The evolution
superoperator for the one-exciton band with the time dependent
Redfield relaxation tensor has a form of a time ordered exponential
\begin{equation}\label{eq10}
    \mathcal{U}(t)=\exp\Big{\{}-\frac{i}{\hbar}\int^t_0d\tau
    R(\tau)\Big{\}}~.
\end{equation}
However, with our ansatz, Eq. \ref{eq09} for the Redfield tensor, we can
evaluate the propagator as an ordinary exponential
\begin{equation}\label{eq11}
    \mathcal{U}(t)=\exp\Big{\{}-\frac{i}{\hbar}R_{\infty}\frac{e^{-\gamma t}+\gamma
    t-1}{\gamma}\Big{\}}~,
\end{equation}
which greatly speeds up the calculation. For the calculations of the
rates we use a single overdamped Brownian oscillator with $\gamma
 = 0.01~$fs$^{-1}$.

\begin{figure}[h]
\includegraphics[width=0.65\textwidth]{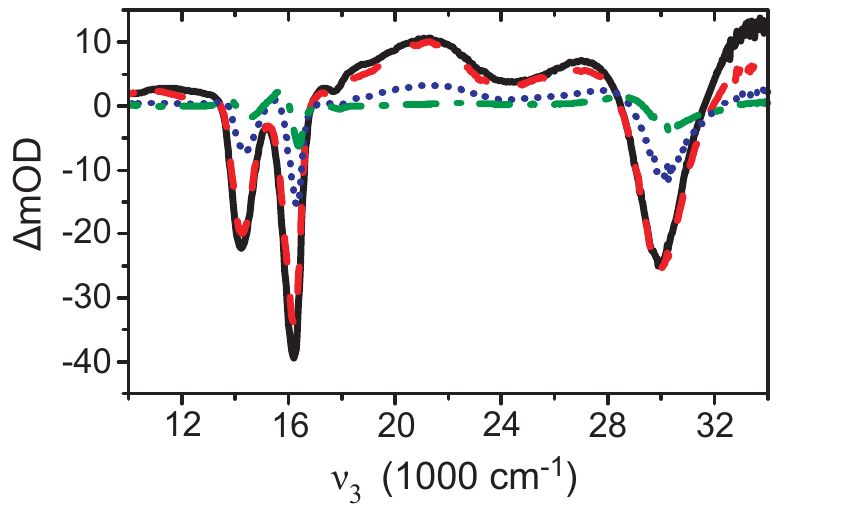}
\caption{\label{figS04} Pump-probe spectra for [LuPc$_2$]$^-$TBA$^+$
at $t_2=$ 100~fs (black solid), $t_2=$ 500~fs (red dash), $t_2=$
5~ps (blue dot), and $t_2=$ 15~ps (olive dash-dot).}
\end{figure}

\begin{figure}[h]
\includegraphics[width=0.65\textwidth]{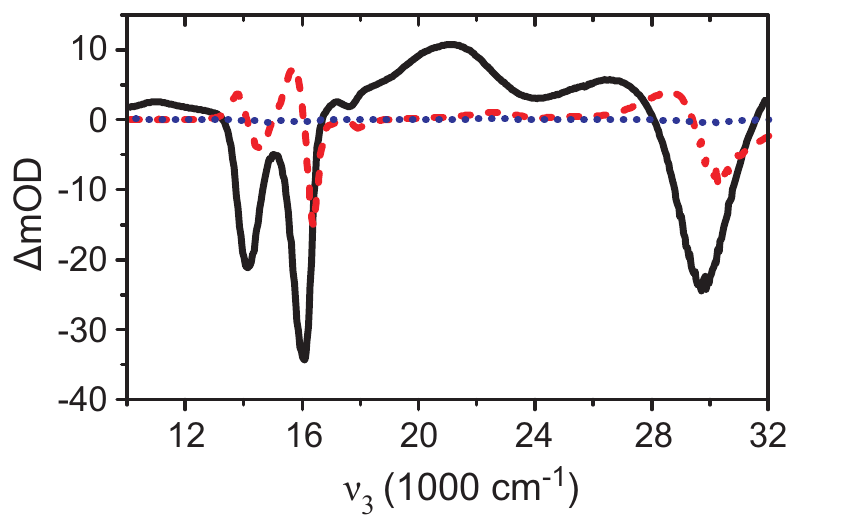}
\caption{\label{figS05} Decay associated difference spectra for
[LuPc$_2$]$^-$TBA$^+$. The three time constants were 3.8~ps (black
solid), 15.8~ps (red dash), and 1~ns (blue dot).}
\end{figure}

\section{Summary of the Pump-Probe Results}
The pump-probe results for [LuPc$_2$]$^-$TBA$^+$ presented in this
study reveal similar results to those previously published by Prall
\textit{et al.} for the 13000-20000 cm$^{-1}$ range
\cite{Prall2005}. Extending the probe range to cover the UV and NIR
regions,\cite{Megerle2009,Herrmann2011} we can conclude that the
dynamics in the B-band region is very similar to the Q-region. Just
like in the previous study, we find that the spectra for $t_2 >$
100~fs are independent of which band we pump. The transient spectra
in Fig. \ref{figS04} show how the Q- and B-band regions
(14000-16000~cm$^{-1}$ and 30000~cm$^{-1}$ respectively) GSB signals
decay on a very similar timescale.

We fitted the transient spectra to three exponentials and the decay
associated difference spectra are shown in Fig. \ref{figS05}. The first
component, with a 3.8~ps lifetime, has contributions from the GSB
regions as well as ESA both in the NIR and VIS region. This also
gives us the lifetime of the lowest state of the single exciton
manifold. This excited state decays into a hot ground state, as
evident by the dispersive lineshape of the second 15.8~ps component,
representing cooling in the ground state. Similar conclusions were
drawn by Prall \textit{et al.} \cite{Prall2005}, but can be
re-enforced by the observation of the same type of features in both
the Q- and -B band regions. An additional very slow component was
needed to account for residual signals. However, as can be seen from
Fig. \ref{figS05}, the amplitude of this component is very low.

%